\documentclass[5p,times]{elsarticle}
\journal{Astronomy and Computing}

\usepackage{dblfloatfix}
\usepackage{natbib}
\usepackage{amsmath}
\usepackage{amssymb}
\usepackage[breaklinks,colorlinks]{hyperref}
\usepackage{tikz}
\usepackage{units}
\usepackage{url}
\usepackage{dsfont} % For \mathds

\bibpunct{(}{)}{;}{a}{}{,}

\newcommand{\polycomp}{\texttt{polycomp}}
\newcommand{\libpolycomp}{\texttt{libpolycomp}}\newcommand{\Planck}{\textit{Planck}}

\begin{document}

\title{\textsc{Polycomp}: Efficient and configurable compression of astronomical timelines}
\ead{maurizio.tomasi@unimi.it}
\author{M.~Tomasi}
\address{Dipartimento di Fisica, Universit\`a degli Studi di Milano (Italy).}

\begin{abstract}
This paper describes the implementation of \polycomp, a open-sourced, publicly available program for compressing one-dimensional data series in tabular format. The program is particularly suited for compressing smooth, noiseless streams of data like pointing information, as one of the algorithms it implements applies a combination of least squares polynomial fitting and discrete Chebyshev transforms that is able to achieve a compression ratio $C_r$ up to $\approx 40$ in the examples discussed in this work. This performance comes at the expense of a loss of information, whose upper bound is configured by the user. I show two areas in which the usage of \polycomp{} is interesting. In the first example, I compress the ephemeris table of an astronomical object (Ganymede), obtaining $C_r \approx 20$, with a compression error on the $x, y, z$ coordinates smaller than $1\,\mathrm{m}$. In the second example, I compress the publicly available timelines recorded by the Low Frequency Instrument (LFI), an array of microwave radiometers onboard the ESA \Planck{} spacecraft. The compression reduces the needed storage from $\sim 6.5\,\unit{TB}$ to $\approx 0.75\,\unit{TB}$ ($C_r \approx 9$), thus making them small enough to be kept in a portable hard drive.
\end{abstract}

\begin{keyword}
coding theory \sep Information systems~Data compression \sep methods: numerical

\PACS 
\end{keyword}

\maketitle

\section{Introduction}

It is increasingly common for astronomers to deal with huge datasets, either produced by means of simulations or measured by instruments. This situation has caused a sharp rise in the demand of disk storage for preserving measurements and simulations in digital format. A telling example is the amount of storage required by three space instruments devoted to the characterization of the CMB anisotropies. The COBE/DMR, which took its measurements in the years 1989--1993, produced less than 8\,GB of time-ordered data\footnote{\url{http://lambda.gsfc.nasa.gov/product/cobe/dmr_prod_table.cfm}.}; the WMAP spacecraft produced roughly 200\,GB of data\footnote{\url{http://lambda.gsfc.nasa.gov/product/map/current/m_products.cfm}.} in the years 2001--2010; lastly, the recently released \Planck\ timelines require $\sim 30$\,TB \citep{planck2014-ES} of disk space, of which 7\,TB are needed for the timelines of the Low Frequency Instrument (LFI), which is one of the examples considered in this paper. In the future, storage requirements for astronomical experiments are going to be even more demanding \citep{Norris.2010.data.challenges.for.future.radio.telescopes,2011arXiv1110.3193L,doi:10.1117/12.2055539,2015arXiv151207914J}. Such huge quantities of data call for efficient data compression algorithms, in order to reduce the requirements in data storage and potentially to speed-up computations by avoiding I/O-related bottlenecks.

In this paper I discuss the implementation of a C library, \libpolycomp\footnote{\url{http://ascl.net/code/v/1373}.}, as well as an open-source Python program, \polycomp\footnote{\url{https://github.com/ziotom78/polycomp}.}, which interfaces to the library through bindings written in Cython\footnote{\url{http://cython.org/}.}. The library implements a number of widely-known compression schemes. Such compression algorithms are applicable to some kinds of one-dimensional timelines that are commonly found in astronomy and cosmology. In particular, one of the algorithms is a new variant of two well-known techniques, polynomial fitting \citep[e.g.,][]{Ohtani2013} and selective filtering of discrete Fourier transforms. This algorithm is especially well suited for smooth, slowly varying series of data with negligible noise, like pointing information. It is a lossy algorithm where the upper bound on the compression errors is tunable by the user.

I discuss the application of \polycomp{} to the compression of two datasets: the ephemeris table for an astronomical object (Ganymede), and the timelines of the Low Frequency Instrument (LFI), an array of microwave radiometers for the measurement of anisotropies of the Cosmic Microwave Background on-board the \Planck{} spacecraft. The latter example is extremely interesting, as the raw data amount to roughly 6.5\,TB and can be compressed by \polycomp{} down to less than 1\,TB. Finally, I estimate how much the \polycomp{} compression impacts a few examples of LFI data analysis, both in terms of compression error and decompression speed.

\subsection{Basic definitions}

In this section, I revise a few standard definitions used in the theory of compressors, for the sake of readers not accustomed with the terminology. A \emph{compression algorithm} takes as input a sequence $\{d_i\}$ of $N$ numbers (or symbols), each $n_\mathrm{bits}^\mathrm{in}$ bits wide, and produces a sequence of $M$ numbers $\{c_i\}$, each $n_\mathrm{bits}^\mathrm{out}$ bits wide, such that $n_\mathrm{bits}^\mathrm{out}\,M < n_\mathrm{bits}^\mathrm{in}\,N$ on average. There must also be an inverse transformation that is able to recover the $N$ input elements $\{d_i\}$ from $\{c_i\}$. The efficiency of the compression\footnote{Obviously, it is not possible to produce a compression algorithm that satisfies the condition $C_r > 1$ for \emph{any} input $\{d_i\}$. What we require here is that it exists a non-trivial class of datasets $\{d_i\}$ for which this happens.} is quantified by the \emph{compression ratio}:
\begin{equation}
\label{eq:compressionRatio}
C_r \triangleq \frac{n_\mathrm{bits}^\mathrm{in}\,N}{n_\mathrm{bits}^\mathrm{out}\,M},
\end{equation}
which in the average case should be greater than 1 (the symbol $\triangleq$ denotes a definition). If no exact inverse transformation exists, but some quasi-inverse formula is able to recover $N$ values $\{\tilde d_i\}$ that approximate the input values $\{d_i\}$, the compression is said to be \emph{lossy} and the quality of the approximation is usually characterized by $\epsilon_c$:
\begin{equation}
\label{eq:compressionError}
\epsilon_c \triangleq \max_{i=1\ldots N} \left|d_i - \tilde d_i\right|.
\end{equation}
The goal of a lossy compression scheme is to achieve the maximum $C_r$ while satisfying some \textit{a priori} requirements on $\epsilon_c$.

We use also another definition of compression ratio, which takes into account \emph{all} the data, metadata, and headers that are needed to decompress the output sequence of $M$ values:
\begin{equation}
\label{eq:compressionRatioPc}
C_r^\mathtt{pc} \triangleq \frac{N_\mathrm{bits}^\mathrm{in}}{N_\mathrm{bits}^\mathrm{out}},
\end{equation}
where $N_\mathrm{bits}^\mathrm{in}$ is the overall number of bits needed to encode the input sequence $\{d_i\}$, and $N_\mathrm{bits}^\mathrm{out}$ is the overall number of bits of the output sequence, including any ancillary data structure. In this work I will use either Eq.~\eqref{eq:compressionRatio} or Eq.~\eqref{eq:compressionRatioPc}, depending on the context. Equation~\eqref{eq:compressionRatioPc} will always refer to bitstreams produced by the \polycomp{} program, hence the superscript \texttt{pc}.

\section{Compressing smooth data series}
\label{sec:algorithmDescription}

The \polycomp{} program implements a number of compression schemes to compress one-dimensional tables read from FITS files. The list of compression algorithms currently implemented by \polycomp{} is the following:
\begin{enumerate}
\item Run-Length Encoding (RLE);
\item Quantization;
\item Polynomial compression;
\item Deflate/Lempel-Ziv compression (via the \texttt{zlib} library\footnote{\url{http://www.zlib.net/}.});
\item Burrows-Wheeler compression (via the \texttt{bzip2} library\footnote{\url{http://www.bzip.org/}}).
\end{enumerate}
The \polycomp{} program saves compressed streams into FITS files containing binary HDUs. The program can act both as a compressor or a decompressor.

In the next sections I will discuss how each algorithm has been implemented, and what are the kinds of data streams for which it provides the best results.

\subsection{Run-Length Encoding}
\label{sec:RLE}

\begin{figure}
	\centering
	\begin{tikzpicture}
		[value/.style={rectangle,minimum size=2em,draw},
		 count/.style={rectangle,minimum size=2em,font=\bfseries,fill=gray!50,draw},
		 diff/.style={font=\footnotesize}]
	
	\node at (-5em, 0) {Input:};
	
	\node at (0em, 0) [value] {5};
	\node at (2em, 0) [value] {5};
	\node at (4em, 0) [value] {5};
	\node at (6em, 0) [value] {5};
	\node at (8.5em, 0) [value] {9};
	\node at (10.5em, 0) [value] {9};
	\node at (12.5em, 0) [value] {9};
	
	\begin{scope}[yshift=-3em]
		\node at (-5em, 0) {Output:};

		\node at (0em, 0) [count] {4};
		\node at (2em, 0) [value] {5};
		\node at (4.5em, 0) [count] {3};
		\node at (6.5em, 0) [value] {9};
	\end{scope}

	\begin{scope}[yshift=-7.5em]
		\node at (-5em, 0) {Input:};
	
		\node (FirstSampleIn) at (0em, 0) [value] {14};
		\node at (2em, 0) [value] {17};
		\node at (4em, 0) [value] {20};
		\node at (6em, 0) [value] {23};
		\node at (8em, 0) [value] {27};
		\node at (10em, 0) [value] {30};
		\node at (12em, 0) [value] {33};
		\node at (14em, 0) [value] {36};
		\node at (16em, 0) [value] {39};
		
		\node at (1em, -1.5em) [diff] {3};
		\node at (3em, -1.5em) [diff] {3};
		\node at (5em, -1.5em) [diff] {3};
		\node at (7em, -1.5em) [diff] {4};
		\node at (9em, -1.5em) [diff] {3};
		\node at (11em, -1.5em) [diff] {3};
		\node at (13em, -1.5em) [diff] {3};
		\node at (15em, -1.5em) [diff] {3};

	\end{scope}
	
	\begin{scope}[yshift=-11em]
		\node at (-5em, 0) {Output:};

		\node (FirstSampleOut) at (0em, 0) [value] {14};
		\node at (2.5em, 0) [count] {3};
		\node at (4.5em, 0) [value] {3};
		\node at (7em, 0) [count] {1};
		\node at (9em, 0) [value] {4};
		\node at (11.5em, 0) [count] {4};
		\node at (13.5em, 0) [value] {3};
	\end{scope}
	
	\draw[->] (FirstSampleIn) -- (FirstSampleOut);
	
	\end{tikzpicture}
	\caption{\label{fig:RLECompression} \emph{Top:} Example of RLE compression applied to an input sequence of 7 values. The output consists of 4 values grouped in two pairs: the first element in each pair (bold text over a gray background) is the repeat count, the second element the value to be repeated. \emph{Bottom:} The differenced RLE algorithm implemented in \polycomp{} stores the value of the first element in the output (in this example, 14), and then it applies a plain RLE to the consecutive differences between adjacent values in the input stream (shown in the drawing as small numbers below the input values).}
\end{figure}
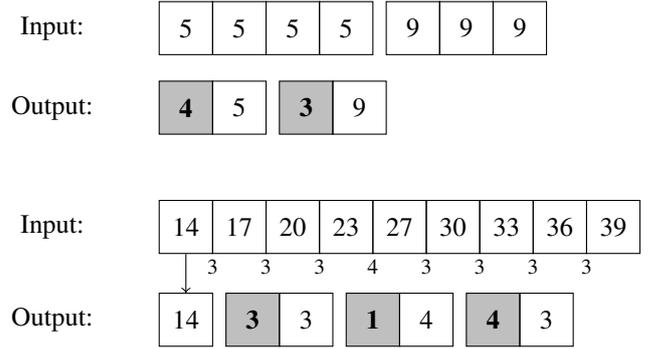

This widely-used algorithm \citep[see e.g.][]{Salomon:2006:DCC:1196474} achieves good compression ratios for input data containing long sequences of repeated values. It detects such sequences and writes pairs of repeat counts and values to the output stream. No information is lost in the process, but if there are not enough repetitions in the input sequence, the compression ratio $C_r$ defined in Eq.~\eqref{eq:compressionRatio} might be less than 1 (the lower bound is $C_r = 1/2$, if no repetitions at all are present in the input data). This algorithm is useful for compressing data flags in timelines, as they usually make very long sequences of repeated values.

The \polycomp{} program implements also a variant\footnote{The current implementation of \polycomp{} only allows to apply the two RLE algorithms described here on sequences of integer numbers.} of RLE: in this case, the algorithm is not applied to the input data $\{d_j\}_{j=1}^N$, but to the differences $\{\Delta_j = d_{j+1} - d_j\}_{j=1}^{N-1}$. The first value $d_1$, needed to decompress the sequence, is saved at the beginning of the output stream. The typical situation where this kind of compression is useful is for quantities that measure the passing of time, if the sampling frequency is kept constant during the acquisition.

See Fig.~\ref{fig:RLECompression} for an example of the application of both variants of the RLE algorithm.

\subsection{Quantization}

\label{seq:quantization}

\begin{figure}
	\centering
	\begin{tikzpicture}
		[fpvalue/.style={rectangle,minimum size=3em,draw},
         value/.style={rectangle,minimum size=3em,draw},
         bits/.style={rectangle,minimum size=3em,draw,font=\small}]
	
	\node at (-5em, 0) {Input:};
	
	\node at (0em, 0) [fpvalue] {3.06};
	\node at (3em, 0) [fpvalue] {5.31};
	\node at (6em, 0) [fpvalue] {2.25};
	\node at (9em, 0) [fpvalue] {7.92};
	\node at (12em, 0) [fpvalue] {4.86};
	
	\begin{scope}[yshift=-3.5em]
		\node at (-5em, 0) {Scaled input:};
		\node at (-5em, -1em) {($0\ldots 2^5-1$)};

		\node at (0em, 0) [value] {4};
		\node at (3em, 0) [value] {17};
		\node at (6em, 0) [value] {0};
		\node at (9em, 0) [value] {31};
		\node at (12em, 0) [value] {14};
	\end{scope}

	\begin{scope}[yshift=-7em]
		\node at (-5em, 0) {Bits ($n = 5$):};
	
		\node at (0em, 0) [bits] {00100};
		\node at (3em, 0) [bits] {10001};
		\node at (6em, 0) [bits] {00000};
		\node at (9em, 0) [bits] {11111};
		\node at (12em, 0) [bits] {01110};
	\end{scope}
	
	\begin{scope}[yshift=-10.5em]
		\node at (-5em, 0) {After packing:};
		
		\begin{scope}[right]
			\node at (0, 0) {\texttt{00100100}$_2$ = 36};
			\node at (0, -1.5em) {\texttt{01000001}$_2$ = 65};
			\node at (0, -3.0em) {\texttt{11110111}$_2$ = 247};
			\node at (0, -4.5em) {\texttt{0\underline{0000000}}$_2$ = 0};
		\end{scope}
	\end{scope}
	
	\end{tikzpicture}
	\caption{\label{fig:Quantization} Example of quantization and bit-packing. The input sequence $\{d_i\}$ is scaled using Eq.~\protect\eqref{eq:quantizationFormula}. The binary representation of each number (using $n = 5$ bits) is packed into 8-bit numbers. Since the number of bits is $25 = 8 \times 4 + 1$, the last bit is stored in a full 8-bit number, whose last 7 bits (underlined) are set to zero. Decompressing the sequence $(36, 65, 247, 0)$ would yield the numbers $(\sim 2.982, \sim 5.359, 2.25, 7.92, \sim 4.811)$.}
\end{figure}

Quantization is a simple way to reduce the entropy of a sequence of numbers by reducing the precision of the numbers; as described in \citet{Salomon:2006:DCC:1196474}, this can be achieved by means of a rounding operation, possibly associated with a scalar operation. The purpose of the latter is to tune the amount of information that is lost in the process. This technique can be applied in several contexts: for instance, when recording floating-point data from digital instruments, it is often the case that such data have been obtained by means of digital integrators. The number of binary digits used by such integrators is usually smaller than the number used to store floating-point numbers on modern CPUs. Another well-known case where quantization plays an important role is in the JPEG compression \citep{Pennebaker:1992:JSI}, where quantization is used as a pre-processing stage before the application of the Huffman or arithmetic compression to the discrete cosine transform coefficients of the image pixels.

The program \polycomp{} can apply a quantization to a sequence of floating-point numbers. The amount of quantization can be configured by the user by means of the parameter $n$, which is the number of bits that must be used for each sample. The input data $\{d_i\}$ are transformed into a set of integer numbers $\{\tilde d_i\}$ through the following formula:
\begin{equation}
\label{eq:quantizationFormula}
\tilde d_i = \left[(2^n - 1)\,\frac{d_i - \min_k d_k}{\max_k d_k - \min_k d_k}\right],
\end{equation}
where $[\cdot]$ denotes a rounding operation. All the numbers $\{\tilde d_i\}$ are in the interval $[0, 2^n - 1]$ and can therefore be encoded using $n$ bits each. The binary encoding of each number is then packed into a sequence of 8-bit bytes. An example is shown in Fig.~\ref{fig:Quantization}.

Decompression is just a matter of inverting Eq.~\eqref{eq:quantizationFormula}, where the inversion is not exact because of the rounding operation. An upper bound to the discrepancy $\epsilon_i$ caused by the rounding operation can be estimated from Eq.~\eqref{eq:quantizationFormula} and from the fact that $\bigl|[x] - x\bigr| \leq 1/2$:
\begin{equation}
\epsilon_i \triangleq \left|d^\textrm{decompr}_i - d_i\right| \leq \frac{\max_k d_k - \min_k d_k}{2(2^n - 1)},
\end{equation}
where $d^\textrm{decompr}_i$ is the $i$-th sample decompressed from the compressed output, and $d_i$ is the $i$-th sample in the input stream. For the example in Fig.~\ref{fig:Quantization}, the upper bound on $\epsilon_i$ is 0.186.

\subsection{Polynomial compression}
\label{sec:polynomialCompressionDescription}

The \libpolycomp{} library implements a new compression algorithm to compress smooth, noise-free 1-D data series, like pointing information and datasets generated through analytical or semi-analytical models. It is an improvement over traditional compression algorithms based on polynomial approximation \citep[there are countless examples of this technique, e.g.,][]{kizner1967-polynomial-fitting,philips1992-polynomial-compression}, and its most natural domains of application are therefore the same. It takes advantage of two widely used families of compression methods: (1)
approximation of the input data through polynomials of low order; and (2) quantization/truncation of Fourier/wavelet transforms; notable examples of this are the JPEG compression scheme \citep{Pennebaker:1992:JSI}, and the MPEG-1 Audio Layer specification\footnote{\url{http://www.iso.org/iso/iso_catalogue/catalogue_tc/catalogue_detail.htm?csnumber=22412}.}.
These two families of compression schemes have a number of properties in common:
\begin{enumerate}
\item They subdivide the data to be compressed into blocks, and treat each block separately;
\item They build a model for the input data, which is able to approximate them up to some level using considerably less information;
\item The quality of the compression is tunable.
\end{enumerate}
Both families have their own advantages and disadvantages: polynomial fitting can be computationally demanding, especially if the degree of the polynomial is high, but it can interpolate slowly-varying, noise-free data very well. On the other hand, Fourier/wavelet techniques are fast, but if the data to be compressed are too regular, they can produce significantly worse compression ratios than polynomial fitting.

The \libpolycomp{} library implements an algorithm that combines both approaches. After having split the input dataset in subsets, called \textit{chunks}, the program computes a least-square polynomial fit $p(x)$ of the data. In the case where $p(x)$ does not allow to reconstruct the input data with the desired accuracy, \polycomp{} computes a Chebyshev transform of the fit residuals, and it saves only those coefficients which allow to reconstruct the input data with the desired precision. I call this algorithm \textit{polynomial compression}.

In the following paragraphs, I use the notation found in \cite{Briggs.Henson.1995.dft} for Chebyshev transforms:
\begin{eqnarray}
\label{eq:directChebyshevTransform}
F_k &=& \frac2{N - 1}\sum_{n=1}^N{}'' g_n\,\cos\left(\frac{\pi(n-1)(k-1)}{N - 1}\right),\\
\label{eq:inverseChebyshevTransform}
g_k &=& \sum_{n=1}^N{}'' F_n\,\cos\left(\frac{\pi(n-1)(k-1)}{N - 1}\right),
\end{eqnarray}
where
\begin{equation}
\sum_{n=1}^N{}'' x_n \triangleq \frac{x_1}2 + \sum_{n=2}^{N-1} x_n + \frac{x_N}2.
\end{equation}

\begin{figure}
	\centering
	\includegraphics[width=\columnwidth]{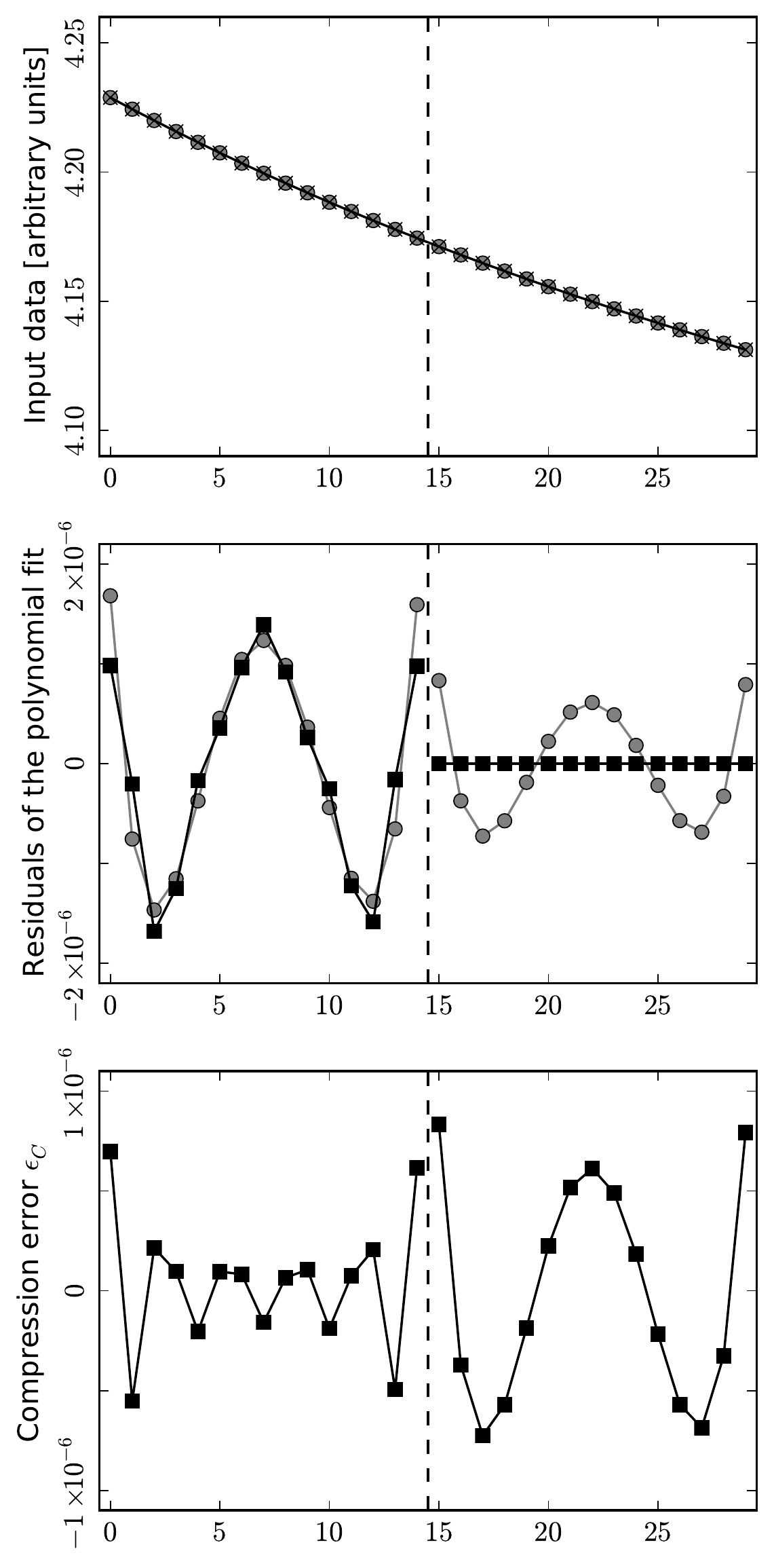}
	\caption{\label{fig:algorithm} Example of an application of the polynomial compression algorithm. \emph{Top:} The $N=30$ data $\{d_j\}$ to be compressed are shown as gray circles. The parameters of the compression are: the maximum compression error $\epsilon_C = 10^{-6}$, the degree $\deg p(x) = 4$ of the interpolation, and the chunk size $N_\mathrm{chunk}=15$. The points are approximated by $p_j$ (crosses), which is the value at $j$ of the polynomial $p(x)$ (one per chunk) which best fits the input data. \emph{Middle:} Plot of $r_j = d_j - p_j$ (gray circles), the discrepancy between the polynomial interpolation and the datum itself. The set of values $\{r_j\}$ in each chunk is approximated by a filtered Chebyshev transform $\{\tilde r_j\}$ (black squares). In the first chunk (\textit{left}), only 9  of 15 Chebyshev coefficients were kept. In the second chunk (\textit{right}), since $\max\left|r_j\right| < \epsilon_C$, no Chebyshev transform was computed and $\tilde r_j = 0\,\forall j$. \emph{Bottom:} Compression error $\epsilon_j = d_j - p_j - \tilde r_j$. This is the difference between the gray circles and the black squares in the middle panel.}
\end{figure}

To apply this algorithm, \polycomp{} requires the following inputs:
\begin{enumerate}
\item a set of $N$ points $\left\{d_j\right\}_{j=1}^N$;
\item a predefined degree for the polynomial $p(x)$, indicated with $\deg p(x)$;
\item an upper bound $\epsilon_c$ for the compression error, as defined in Eq.~\eqref{eq:compressionError};
\item A subdivision of the sequence of $N$ samples $\{d_i\}$ into subsets $D_k$, called \emph{chunks}, of consecutive elements. It is not mandatory for the chunks to have the same number of elements; however, for simplicity\footnote{Such assumption might be generalized in future versions of the program.} \polycomp{} splits the input dataset in a number of chunks with the same number of elements each, with the possible exception of the last one.
\end{enumerate}
The polynomial compression algorithm works as follows:
\begin{enumerate}
\item It splits the data set $\left\{d_j\right\}_{j=1}^N$ in chunks, and apply the following passages to each of them. I indicate the number of elements in the chunk with $N_\text{chunk}$.
\item It calculates the coefficients of the least-square fitting polynomial $p(x)$ which fits the points $(j, d_j)_{j=1}^{N_\text{chunk}}$. This polynomial is used to define the values $\left\{p_j\right\}_{j=1}^N$, where $p_j \triangleq p(j)$.

\item\label{step:calculateResiduals} It calculates the residuals $r_j$ between $p(x)$ and $d_j$:
\begin{equation}
\label{eq:residuals}
r_j = d_j - p_j, \quad j = 1 \ldots N_\text{chunk}.
\end{equation}
If $\max_j \left|r_j\right| \leq \epsilon_c$, then the knowledge of the coefficients of $p(x)$ is enough to reconstruct all the samples in the chunk with the desired accuracy. According to Eq.~\eqref{eq:compressionRatio}, the compression ratio for this chunk is
\begin{equation}
\label{eq:polyComprRatio}
C_r = \frac{N_\text{chunk}}{\deg p(x) + 1},
\end{equation}
if $n_\mathrm{bits}^\mathrm{in} = n_\mathrm{bits}^\mathrm{out}$.

\item\label{step:ChebyshevResiduals} If $\max_j \left|r_j\right| > \epsilon_c$, then \polycomp{} decomposes the set of numbers $\{r_j\}_{j=1}^{N_\text{chunk}}$ defined in Eq.~\eqref{eq:residuals} using the direct Chebyshev transform in Eq~\eqref{eq:directChebyshevTransform}. The set of transformed numbers $\left\{R_j\right\}_{j=1}^N$ is then filtered so that only the $N_t$ elements with the greatest absolute value are kept, while the others are set to zero and not saved in the compressed stream. The number $N_t$ is chosen as the smallest able to ensure that the following two conditions hold:
\begin{align}
\label{eq:NtDefinitionI}
1 \leq N_t < N_\mathrm{chunk} - \deg p(x) - 1 - \lceil N_t / 8 \rceil, \\
\label{eq:NtDefinitionII}
\max_j \left|d_j - \bigl(p(j) + \tilde{r}_j\bigr)\right| \leq \epsilon_c,
\end{align}
where $\tilde{r}_j$ is the result of the inverse Chebyshev transform applied to the filtered list of $N_t$ elements of $\{R_j\}$, with the filtered positions filled with zeroes. Condition \eqref{eq:NtDefinitionI} ensures that $C_r > 1$, as it is shown below. In this case, the $N_t$ filtered coefficients of the Chebyshev transform must be saved alongside the $\deg p(x) + 1$ coefficients of polynomial $p(x)$, together with a bitmask that allows to determine their indices. The compression ratio in this case is
\begin{equation}
\label{eq:chebyComprRatio}
C_r = \frac{N_\text{chunk}}{\deg p(x) + 1 + N_t + \lceil N_t / 8 \rceil},
\end{equation}
where the term $\lceil N_t/8 \rceil$ quantifies the storage for the bitmask.

\item\label{step:uncompressedChunk} From Eq.~\eqref{eq:chebyComprRatio}, for those chunks where $\deg p(x) + 1 + N_t > N_\text{chunk}$, \polycomp{} saves the set of values $\{d_j\}$ in uncompressed form.
\end{enumerate}
Figure~\ref{fig:algorithm} illustrates the application of this compression scheme to some test data.

The decompression is straightforward:
\begin{enumerate}
\item Read the coefficients of the polynomial $p(x)$ for the first chunk and computes the set of points $\left\{ p_j \right\}_{j = 1}^{N_\mathrm{chunk}}$;
\item If no Chebyshev coefficients are available, the decompressed data are $\left\{ p_j \right\}_{j = 1}^{N_\mathrm{chunk}}$;
\item If $N_t$ Chebyshev coefficients $\left\{R_j\right\}_{j=1}^{N_t}$ are available, produce a sequence of $N_\mathrm{chunk}$ values by filling empty positions with zeroes.
\item Compute $\left\{ r_j \right\}_{j = 1}^{N_\mathrm{chunk}}$ using the inverse Chebyshev transform formula (Eq.~\ref{eq:inverseChebyshevTransform}). The decompressed data for this chunk are
\begin{equation}
\tilde d_j = p_j + r_j.
\end{equation}
\item Iterate over the remaining chunks.
\end{enumerate}
Decompression is faster than compression, as there is no need to compute a linear square fit to find the polynomial $p(x)$: the most computationally intensive operations are the evaluation of the polynomial $p(x)$ at $N_\mathrm{chunk}$ points and (when necessary) the computation of the inverse Chebyshev transform, which is done using the FFTW\ 3 (Fastest Fourier Transform in the West) library \citep{Frigo:1999:FFT:301618.301661}.

The values $\deg p(x)$ and $N_\textit{chunk}$ are input parameters for the compressor and need to be properly tuned, in order to produce the desired compression ratio. Their optimization can be tricky, as a number of factors must be considered in choosing them:
\begin{enumerate}
\item Generally, the larger $\deg p(x)$ and $N_\textit{chunk}$, the better the compression ratio.
\item Large numbers for $\deg p(x)$ can produce round-off errors. Such errors are detected by \polycomp{}, but they force the program to degrade the compression ratio by saving more and more Chebyshev coefficients in order to correct the interpolation.
\item Large values of $N_\textit{chunk}$ increase significantly the time required for the polynomial fitting and the direct/inverse Chebyshev transforms.
\end{enumerate}
There are several ways to tackle the problem of tuning a lossy compression algorithm. They depend on the nature of the data and the kind of analyses that are expected to be performed on the data themselves. In some cases, it is possible to derive an analytical model that can predict the best values to be used for the parameters. For instance, \citet{Shamir2005} propose a lossy compression algorithm for astronomical images used for photometry, and it provides a set of equations to quantify the impact of the loss of information to quantities commonly used in photometric analyses. If the development of a theoretical model for the compression is too complex, the most common approach is to estimate the error induced on the results of the data analysis when compressed data are used instead of the original uncompressed ones. For a few examples of the latter approach, see e.g., \citet{Vohl2015,Loeptien2016}.

Considering the broad target of this paper, it would be too impractical to provide analytical models to forecast the impact of compression errors to \emph{every} conceivable scientific product obtained using data compressed with \polycomp. However, the program provides two tools which can ease the choice of the best compression parameters for polynomial compression:
\begin{enumerate}
\item A slow optimization mode, where polycomp tries a set of pairs $(\deg p, N_\text{chunk})$ provided by the user and picks the one with the best compression ratio;
\item A fast optimization mode, where polycomp requires a pair of values $(\deg p, N_\text{chunk})$ as a starting point, and it uses the algorithm proposed by \cite{Nelder01011965} to find the configuration with the best $C_r$.
\end{enumerate}
In both cases, the upper bound on the compression error $\epsilon_c$ must be passed to the optimizer as an input. In Sect.~\ref{sec:compressionPerformance}, I will present some examples of this optimization, and I will discuss a few important caveats to be kept in mind when optimizing the polynomial compression. A few more technical details about \libpolycomp{} are provided in \ref{sec:Polycomp}.

\subsection{Quantifying the relative importance of the polynomial fitting and of the Chebyshev transform}
\label{sec:FitChebyshevRoles}

Depending on the kind of data to compress and on the compression parameters $\deg p(x)$ and $N_\mathrm{chunk}$, the role of Step~\ref{step:ChebyshevResiduals} in the polynomial compression algorithm (Sect.~\ref{sec:polynomialCompressionDescription}) might be of greater or lesser importance to determine the overall compression ratio. We can consider two extreme cases: (1) the required $\epsilon_c$ is so large that Step~\ref{step:ChebyshevResiduals} is never applied; (2) the required $\epsilon_c$ is so small that all chunks must be saved uncompressed (Step~\ref{step:uncompressedChunk}).

A practical way to quantify the importance of the Chebyshev transform in the polynomial compression is to compare the compression ratio with the one achieved using a simpler algorithm which skips Step~\ref{step:ChebyshevResiduals} completely. In the latter algorithm, if the residuals calculated in Step~\ref{step:calculateResiduals} are too large, the chunk is always saved in uncompressed form, and no Chebyshev transform is ever calculated. I call this algorithm \emph{simple polynomial compression}. The library \libpolycomp{} implements the simple compression algorithm as well. In Sect.~\ref{sec:compressionPerformance}, I use this feature to assess the importance of the Chebyshev transform step in the examples considered in this paper.

\subsection{Other compression algorithms}
\label{sec:otherCompressionAlgorithms}

The compression algorithms presented so far are specialized for very particular kinds of datasets. The \polycomp{} program can interface to two widely used, general-purpose compression libraries in those cases where none of the algorithms described above are suitable:
\begin{enumerate}
\item The \texttt{zlib} library, which implements a variant of the LZ77 algorithm called DEFLATE\footnote{\url{http://www.zlib.net/feldspar.html}.};
\item The \texttt{bzip2} library, which implements a combination of the Burrows-Wheeler algorithm and Huffman coding.
\end{enumerate}

\section{Compression performance}
\label{sec:compressionPerformance}

\begin{figure}
    \centering
    \includegraphics[width=\columnwidth]{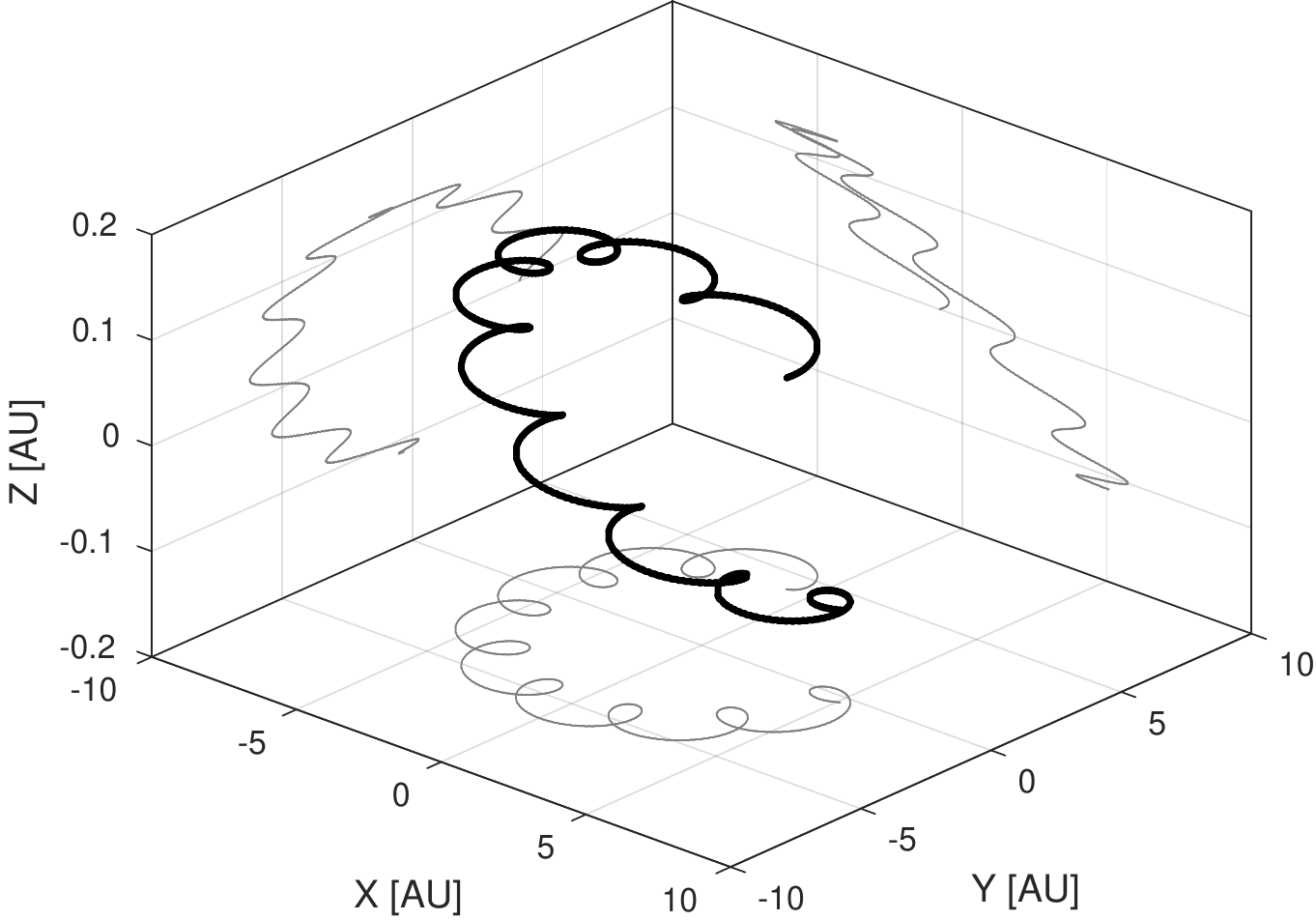}
    \caption{\label{fig:GanymedeOrbit} Ephemeris data for Ganymede, from January 1st, 2002 to December, 31st 2010. The X and Y axis lie on the Ecliptic plane, and have the same size. The Z axis is not in scale with the X and Y axis.}
\end{figure}

In this section I will discuss two applications of the polynomial compression algorithm described in Sect.~\ref{sec:algorithmDescription}, and I compare its performances with other algorithms.

\subsection{Ephemerides}
\label{sec:ephemerides}

I have used \polycomp{} to compress ephemeris data for Ganymede, one of Jupiter's moons. Ephemeris tables are a natural input for polynomial compression, as they are usually smooth in nature (provided that the sampling frequency is not too low). The trajectory of Ganymede in the 3-D space as seen from an observer located in Milan (longitude 9.1912$^\circ$, latitude 45.4662$^\circ$, altitude 147\,m) is shown in Fig.~\ref{fig:GanymedeOrbit}. I consider the interval of time spanning the interval since January, 1st 2002 till December, 31st 2010. I obtained the ephemeris table using JPL's \texttt{HORIZONS} system\footnote{\url{http://ssd.jpl.nasa.gov/?ephemerides}.}. The dataset used for this analysis contains the time (a Julian Date) and the position $(x, y, z)$, measured in AU. These quantities are sampled every 10\,min. I have saved such data into a FITS file containing one binary HDU with four 64-bit floating-point columns. The file contains 473\,328 rows, and its size is 14.45\,MB. It is not easily compressed by standard tools like \texttt{gzip} and \texttt{bzip2}: the compression ratio in these cases is 1.31 and 1.28, respectively, using the \texttt{-9} command-line switch to force both programs to achieve the best possible compression.

To compress the dataset using \polycomp, I have chosen polynomial compression for all the four columns. Since this is a lossy compression, it is necessary to set the upper bound on the compression error (Eq.~\ref{eq:compressionError}). I used $\epsilon_c = 1.16\times 10^{-4}$ for the Julian time, corresponding to an error of the order of 10\,s, and $\epsilon_c = 6.6845871\times 10^{-12}\,\mathrm{AU} = 1\,\mathrm{m}$ for each of the three coordinates $x$, $y$, and $z$. The bounds for $x$, $y$, and $z$ are probably stricter\footnote{The precision of \texttt{HORIZONS} estimates is time-dependent; according to the documentation, ``\textit{uncertainties in major planet ephemerides range from 10\,cm to 100+\,km}'' (\url{http://ssd.jpl.nasa.gov/?horizons_doc\#limitations}).} than the average precision of \texttt{HORIZONS}' ephemerides: they have been chosen because of their interesting properties in the characterization of the polynomial compression algorithm.

\begin{table}[tbf]
	\centering
	\begin{tabular}{lccccccc}
	& $N_\text{chunk}$& $\deg p(x) + 1$& $C_r^\mathtt{pc}$& $C_r^{\mathtt{pc},\mathrm{simple}}$\\
	\hline
	JD& 50\,000& \hphantom{2}2& 10\,518.40& 10\,518.40\\
	$x$& \hphantom{50\,}360& 23& \hphantom{10\,5}11.53& \hphantom{10\,51}4.86\\
	$y$& \hphantom{50\,}360& 22& \hphantom{10\,5}11.63& \hphantom{10\,51}4.64\\
	$z$& \hphantom{50\,}400& 22& \hphantom{10\,5}13.79& \hphantom{10\,5}13.79\\
	\hline
	\end{tabular}
	\caption{\label{tab:ganymedeOptimization} Compression parameters used for the four datasets in the ephemeris table for Ganymede and the resulting compression ratio. All but the first (JD) have been found by \texttt{polycomp} using the ``slow optimization mode'' described in Sect.~\protect\ref{sec:polynomialCompressionDescription}.}
\end{table}

\begin{figure*}[tbf]
	\centering
	\includegraphics[width=\textwidth]{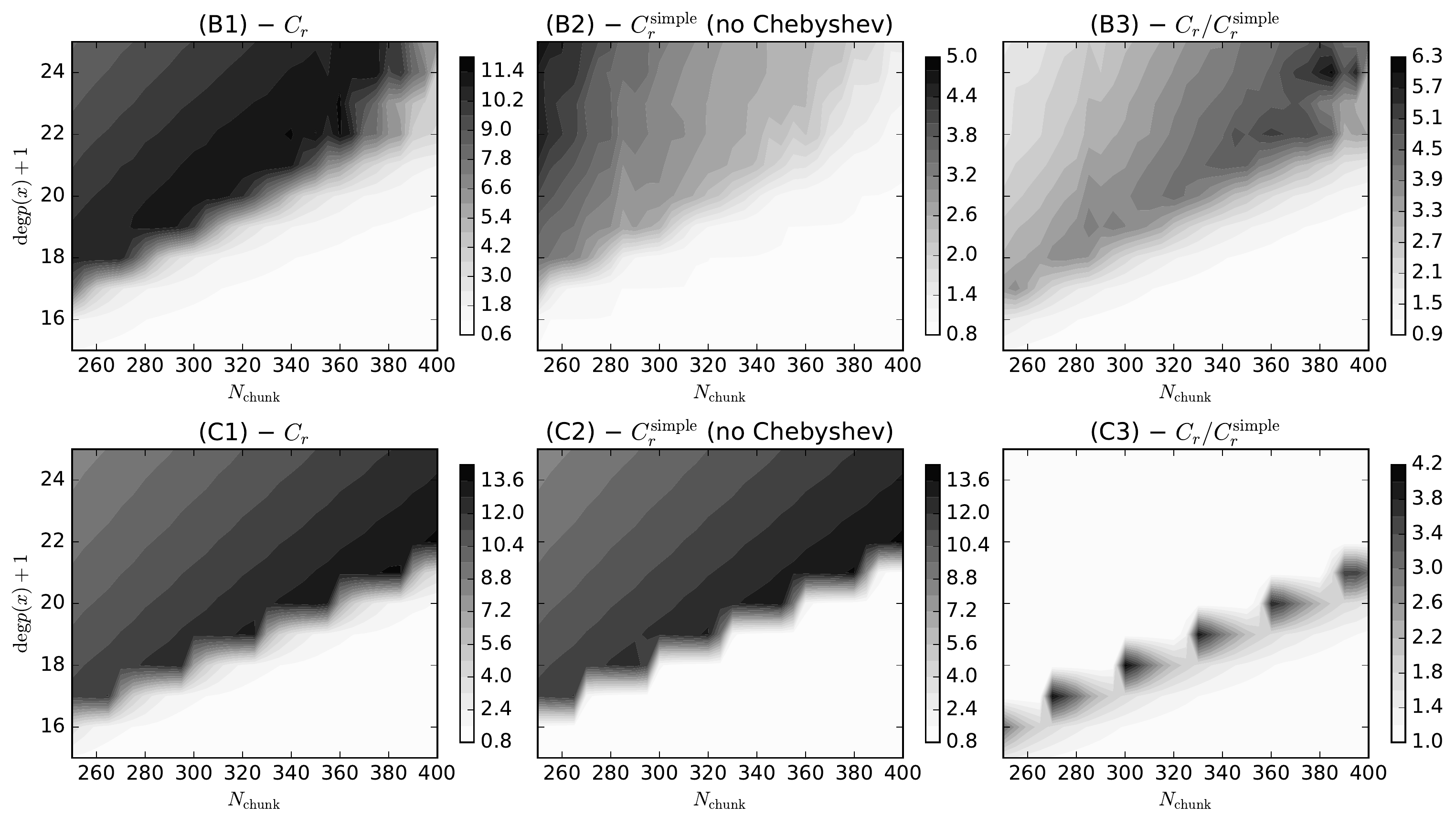}
	\caption{\label{fig:ganymedeOpt} Result of the optimization of the parameters $N_\text{chunk}$ (samples per chunk) and $\deg p(x) + 1)$ (polynomial coefficients) used by the \polycomp{} polynomial compressor for the ephemeris table of Ganymede. \textit{Row A:} the value of $C_r$ when the polynomial compression is applied to the set of $x$ values (A1), the value of $C_r$ when the compression is applied to the same data, but without the Chebyshev transform step (A2), and the ratio between the values shown in panels A1 and A2 (A3). Considering the $y$ values would have produced plots nearly identical to the ones shown here. \textit{Row B:} The same plots as in row A, but using the $z$ coordinate. In this case, the polynomial compression shows no clear advantage over simple compression.}
\end{figure*}

Given the straightforward behaviour of the JD datastream, I avoided the application of a full optimization procedure for the compression parameters and used $\deg p(x) + 1 = 2$, $N_\mathrm{chunk} = 50\,000$. As the data are neatly fitted by a straight line, the compressor does not need to apply any Chebyshev transform, and therefore the full algorithm and the simple compression show the same performance. The expected value for $C_r$ is \begin{equation}
C_r = \frac{N_\mathrm{chunk}}{\deg p(x) + 1} = 25\,000,
\end{equation}
while the measured value for $C_r^\mathtt{pc}$ is about 10\,500.

For $x$, $y$, and $z$, I used the optimization mode provided by \polycomp{} to find the parameters of the compressor that produce the best compression ratio $C_r^\mathtt{pc}$. I explored the region of the parameter space generated by the following numbers:
\begin{align}
\deg p(x) + 1 &\in \{15, 16, 17, \ldots, 25\}, \\
N_\mathrm{chunk} &\in \{250, 255, 260, \ldots, 400\}.
\end{align}
In Fig.~\ref{fig:ganymedeOpt}, I show the result of the exploration of the compression parameter space for a selected number of cases. In the three plots in Row A, I show the difference between the performance of the polynomial compression algorithm versus the simple algorithm when applied to the $x$ dataset. The advantage of the former over the latter is evident; the same behaviour is observed with the data in the dataset $y$, which are not shown here. Row B shows that in the case of the $z$ dataset the Chebyshev step does not give significant advantages over simple compression: as a matter of fact, the best compression ratio in the two cases is the same, as shown in Table~\ref{tab:ganymedeOptimization}. This is a general property of the polynomial compression algorithm: for sufficiently relaxed constraints on $\epsilon_c$, the Chebyshev transform does not provide any advantage over a plain polynomial fitting compression. In the case of the $z$ data, this would have been the case if $\epsilon_c$ had been set to $10\,\mathrm{m}$ instead of $1\,\mathrm{m}$. 

The size of the compressed file is 795.9\,kB, thus $C_r^\mathtt{pc} = 18.6$. This is a substrantial improvement over the simple compression algorithm, which produces a 1681.9\,kB file, with $C_r = 8.8$. On the other hand, if $\epsilon_c = 10\,\mathrm{m}$, then the size shrinks down to 615.9\,kB, with $C_r^\mathtt{pc} = 24.0$: in this case, the simple and polynomial compression schemes produce files of the same size because of the reasons stated above.

\subsection{Planck/LFI pointing information}
\label{sec:LFIPointingCompression}

\begin{figure}[tb]
	\centering
	\includegraphics[width=\columnwidth]{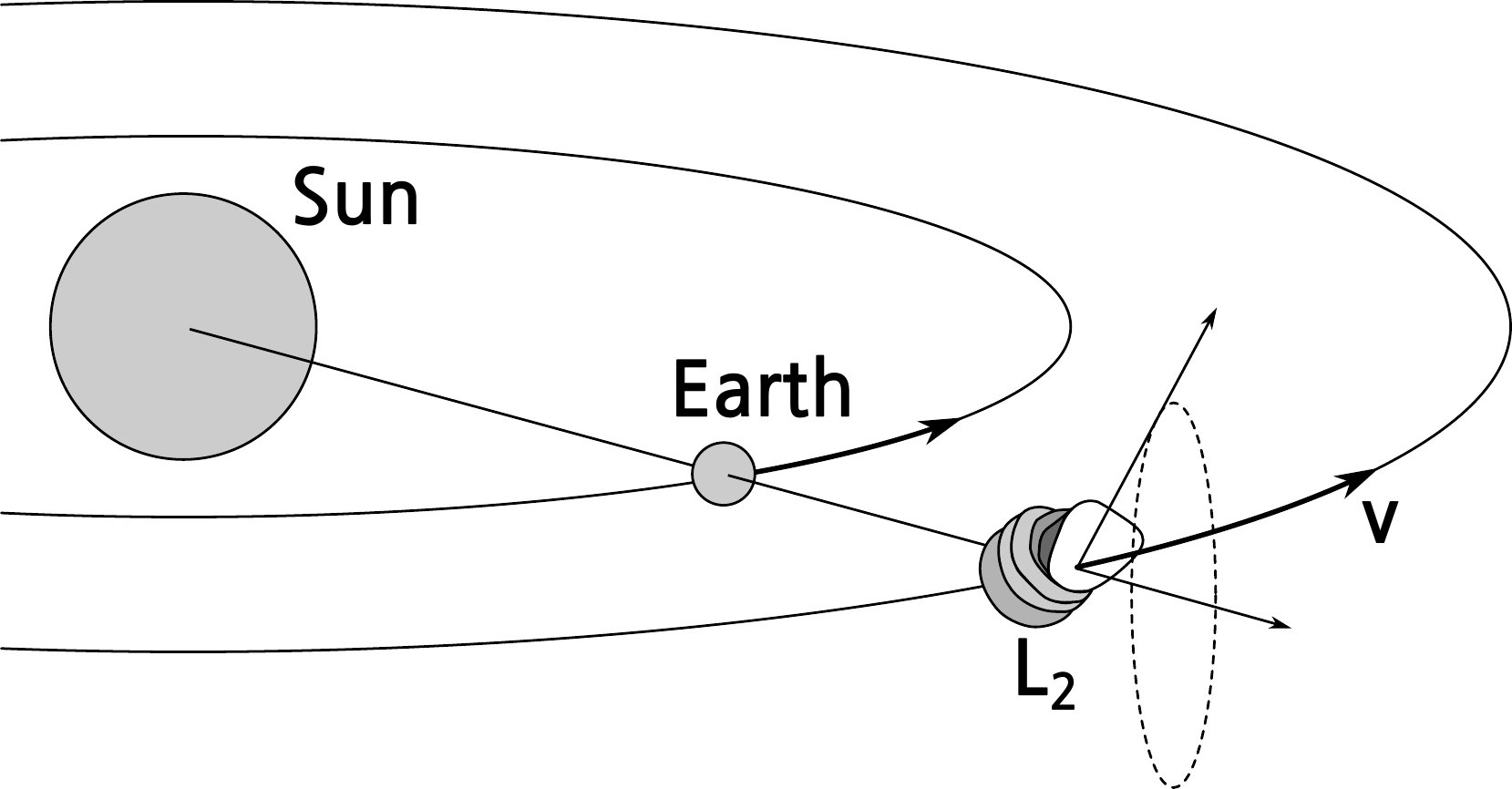}
	\caption{\label{fig:PlanckOrbitSketch} Orbit of \Planck{} around the Sun. The spacecraft orbits around the second Lagrangean point ($L_2$) of the Sun-Earth system, scanning the sky in circles with a pointing direction nearly perpendicular to a spin axis aligned with the Sun-Earth direction. The spacecraft performs one rotation per minute, and every circle is scanned sixty times before the spin axis is tilted by roughly $2.5\,\mathrm{arcmin}$. This scanning strategy produces regular, smooth variations in the pointing direction of the beams.}
\end{figure}

\begin{table*}
    \centering
    \begin{tabular}{ccc}
    Datum & Data format & Compression algorithm \\
    \hline
    On-board time & 64-bit signed integer & Differenced RLE \\
    $\theta$ (colatitude) & 64-bit floating point & Polynomial compression \\
    $\phi$ (longitude) & 64-bit floating point & Polynomial compression \\
    $\psi$ (orientation) & 64-bit floating point & Polynomial compression \\
    Temperature & 64-bit floating point & Quantization \\
    Flags & 16-bit integer & RLE
    \end{tabular}
    \caption{\label{tbl:inputOutputFormat} Format of the data used in the compression of the \Planck/LFI timelines. The ``input format'' refers to the FITS file provided as input to \polycomp, while the ``output format'' and the ``compression algorithm'' specify the kind of the data and the compression scheme used in the file produced by \polycomp. One row of data in the input file requires 44 bytes, of which 24 are used for pointing information ($\theta$, $\phi$, and $\psi$). The ``flags'' column is the combination of the two 8-bit flag columns found in the PLA.}
\end{table*}

In this section, I study the application of the compression algorithm presented in Sect.~\ref{sec:algorithmDescription} to the timelines of the LFI instrument onboard the \Planck{} spacecraft \citep{planck2014-a01}. The LFI (Low Frequency Instrument) is an array of cryogenically cooled HEMT radiometers which observe the sky at three different frequencies: 30, 44, and 70\,GHz . The timelines recorded by the instrument are publicly available through the \Planck\ Legacy Archive\footnote{\url{http://www.cosmos.esa.int/web/planck/pla}.} (PLA); each timeline contains the on-board time, the orientation in the sky of the radiometer's beam, the temperature measured by the radiometer, and a set of quality flags. A relevant fraction of the \Planck{} timelines ($\sim 50\,\%$) contains the information about the pointing direction of the instruments: I refer to such data as \emph{pointing information}, or \emph{pointings}. Pointing information can be encoded either as a set of angles or as length-one vectors, and it can typically take\footnote{Other information recorded in the timelines usually include the timing itself, the scientific datum, and various flags. In the case of the \Planck/LFI timelines available on the \Planck{} Legacy Archive\footnote{\url{http://pla.esac.esa.int}.} (PLA), the timing and the scientific datum take 8 bytes each, while flags require 4 bytes each. On the other side, the pointing information is encoded using three angles $\theta$, $\varphi$, and $\psi$, for a total of 24 bytes. So, 24 bytes out of 44 are needed for the angles. Additional housekeeping timelines like bias currents and instrument temperatures are saved at a much lower sampling rate, and they are not an issue.} 50\,\% of the overall space needed by the timelines. The dependence of the LFI beam orientation on time depends  on the scanning strategy employed by \Planck, which is sketched in Fig.~\ref{fig:PlanckOrbitSketch}.

Pointing information is usually reconstructed using the information about the placement and orientation of the instrument with respect to some reference frame (e.g., the barycentre of the spacecraft), as well as detailed information about the placement of the instrument itself with respect to the center of the reference frame. In some cases, it is enough to combine the line-of-sight vector with the attitude information in order to retrieve the pointing information at any given time\footnote{For instance, this is the approach followed by the WMAP team in publishing the WMAP timelines, see \url{http://lambda.gsfc.nasa.gov/product/map/current/m_products.cfm}.}. However, for instruments with moderate angular resolution and high sensitivity like LFI, a number of systematic effects that need to be taken into account (stellar aberration, variation in the placement of the optically sensitive parts of the instrument due to thermal dilation, etc.) can lead to non-trivial algorithms to reconstruct the pointing information. (See \citet{planck2014-ES}, which details the pipeline used for reconstructing the \Planck{} pointing information.) In such cases, saving the computed pointings alongside the scientific data is the best solution for allowing the scientific community to easily use the timelines. This is the approach followed by the PLA.

I characterize here the application of \libpolycomp's algorithm to the whole set of \Planck/LFI timelines (4 years of data), in terms of the compression ratio $C_r$ (eq.~\ref{eq:compressionRatio}). \ Details about the data formats and the compression algorithms used in the analysis are reported in Table~\ref{tbl:inputOutputFormat}. The layout of the columns used in the input FITS files differs from the layout used by the PLA, as each PLA FITS file contains data acquired by all the radiometers with the same central frequency, and therefore it saves only one copy of the ``On-board time'' column. Since it is usually more handy to analyze each radiometer separately, I split each PLA file into $N$ FITS files, each containing data for one of the $N$ radiometers working at the specified frequency ($N$ is 12, 6, and 4 at 70, 44, and 30\,GHz respectively); this is the same format used internally by the LFI data processing pipeline. The amount of disk space occupied by the $12+6+4=22$ sets of files is 6.5\,TB.

\begin{table}
	\centering
	\begin{tabular}{cccc}
	C.~freq.& $N$& $\nu_\text{samp}$ [Hz]& Radiometer\\
	\hline
	70\,GHz& 12& 78.8& LFI18M\\
	44\,GHz& 6& 46.5& LFI24M\\
	30\,GHz& 4& 32.5& LFI27M\\
	\hline
	\end{tabular}
\end{table}

I have applied the compression algorithms provided by \polycomp{} to the pointing and scientific information of three out of the 22 LFI radiometers, namely LFI18M (70\,GHz), LFI24M (44\,GHz), and LFI27M (30\,GHz). The three radiometers sample the sky temperature with different frequencies: 78.8\,Hz (LFI18M), 46.5\,Hz (LFI24M), and 32.5\,Hz (LFI27M).

\begin{figure*}[tbf]
	\includegraphics[width=\textwidth]{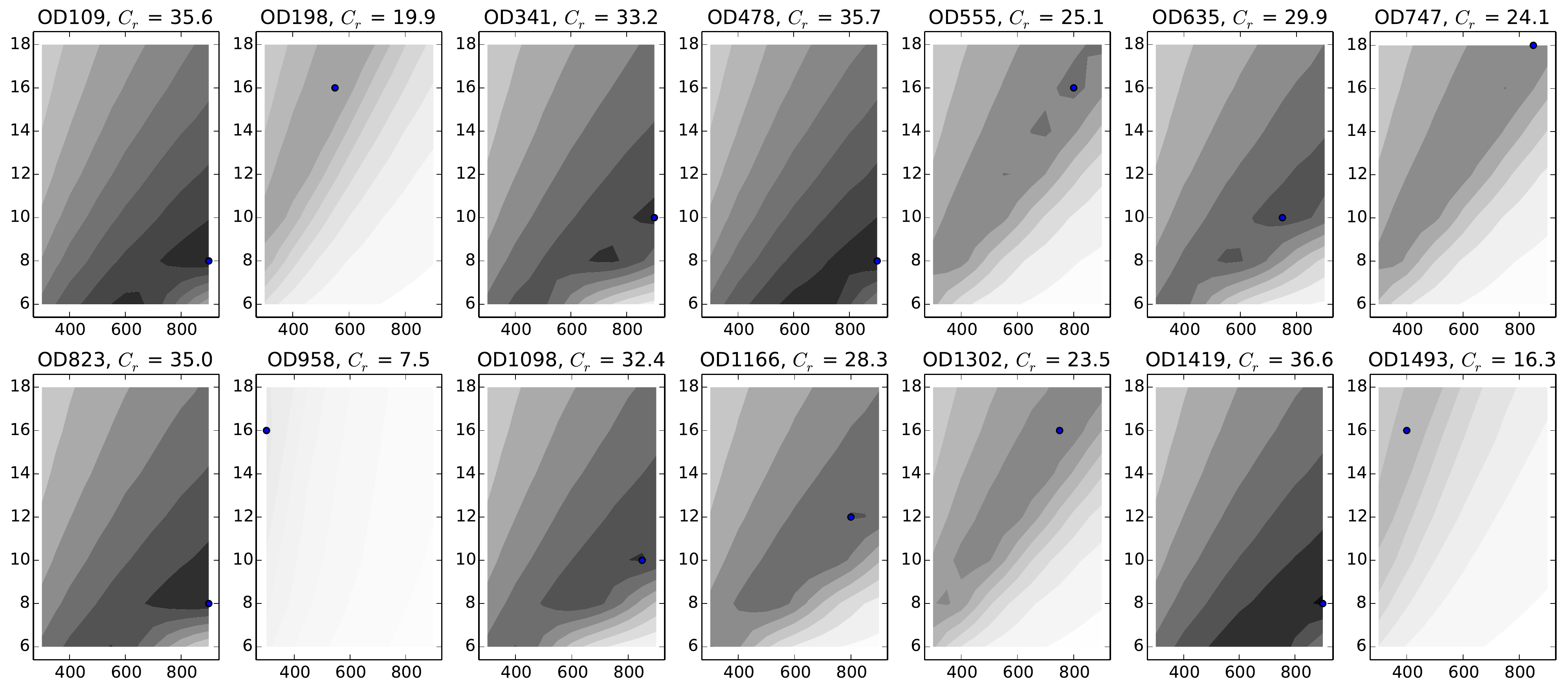}
	\caption{\label{fig:LFIoptimTheta} Optimization of the compression parameters $\deg p(x) + 1$ and $N_\textrm{chunk}$ for the datastream of $\theta$ angles for LFI18M in fourteen operational days (ODs) of data. Blue dots mark the configuration with the highest $C_r$.}
\end{figure*}

\begin{figure}[tb]
	\centering
	\includegraphics[width=\columnwidth]{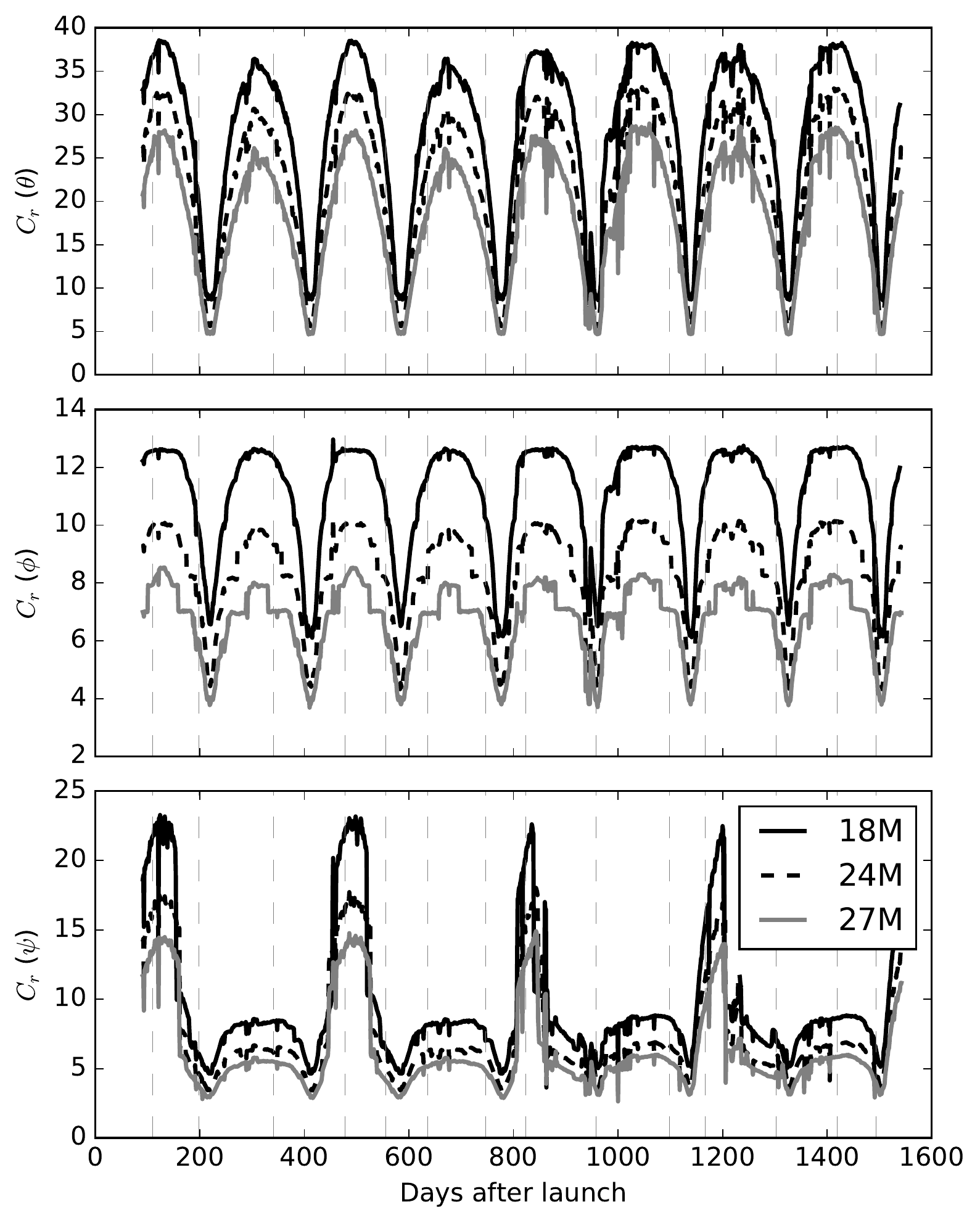}
	\caption{\label{fig:LFIcr} Compression ratio $C_r$ for the daily LFI datasets containing the three pointing angles $\theta$ (colatitude), $\phi$ (longitude), and $\psi$ (orientation of the beam). Three radiometers are considered here: LFI18M (70\,GHz, with a sampling frequency $\nu_s = 78.8\,\mathrm{Hz}$), LFI24M (44\,GHz, $\nu_s = 46.5\,\mathrm{Hz}$), and LFI27M (30\,GHz, $\nu_s = 32.5\,\mathrm{Hz}$). The polynomial compression achieves the best compression ratio for the radiometer with the highest sampling frequency, and the angle showing the best compression performance is the one which varies more slowly, i.e., the colatitude $\theta$. The eight drops in the values of the compression ratio happen after every completion of a sky survey. Vertical dashed lines mark the position of the fourteen operational days (ODs) discussed in Fig.~\protect\ref{fig:LFIoptimTheta}.}
\end{figure}

Choosing the optimal compression strategy for the pointing information is not trivial, as there are 1452 files per radiometer, each corresponding to one Operational Day (OD), and each file has the pointing information encoded in three columns: $\theta$ (Ecliptic colatitude), $\phi$ (Ecliptic latitude), and $\psi$ (orientation of the beam). Therefore, the compressor must be tuned separately for each column. Moreover, since \Planck's scanning strategy varies with time, the best values for the two polynomial compression parameters, $\deg p(x) + 1$ and $N_\mathrm{chunk}$, change with time. For instance, Fig.~\ref{fig:LFIoptimTheta} shows contour plots of the value of $C_r$ as a function of $N_\mathrm{chunk}$ and $\deg p(x) + 1$ for a set of fourteen randomly-chosen days, when \polycomp{} is applied to the set of colatitudes $\theta$ for radiometer LFI18M.

To determine the best parameters for the compressor, I used the optimization function in the \texttt{pypolycomp} Python library to write a script that implemented a two-stage search strategy:
\begin{enumerate}
\item I sampled the parameter space using a wide but coarse grid:
\begin{equation}
\begin{split}
\deg p(x) + 1 &\in \{2, 5, 8, \ldots, 20\}, \\
N_\mathrm{chunk} &\in \{50, 150, 250, \ldots, 950\}.
\end{split}
\end{equation}
To quicken the process, during this process I used only the first two hours of data for each day.
\item After the first run, I ran the optimizer again on a narrower, finely-gridded area around the point with the best $C_r$ that has been found in the previous run, this time using the full, one-day-long dataset.
\end{enumerate}
Fig.~\ref{fig:LFIcr} shows the compression ratio for all the combinations of 1452 operational days (ODs), three angles, and three radiometers considered in this work. There are periodic drops of the compression ratio, and these usually occur at the end of a complete survey of the sky: they are related to quick changes in the asset of the instrument.

\begin{figure}
	\includegraphics[width=\columnwidth]{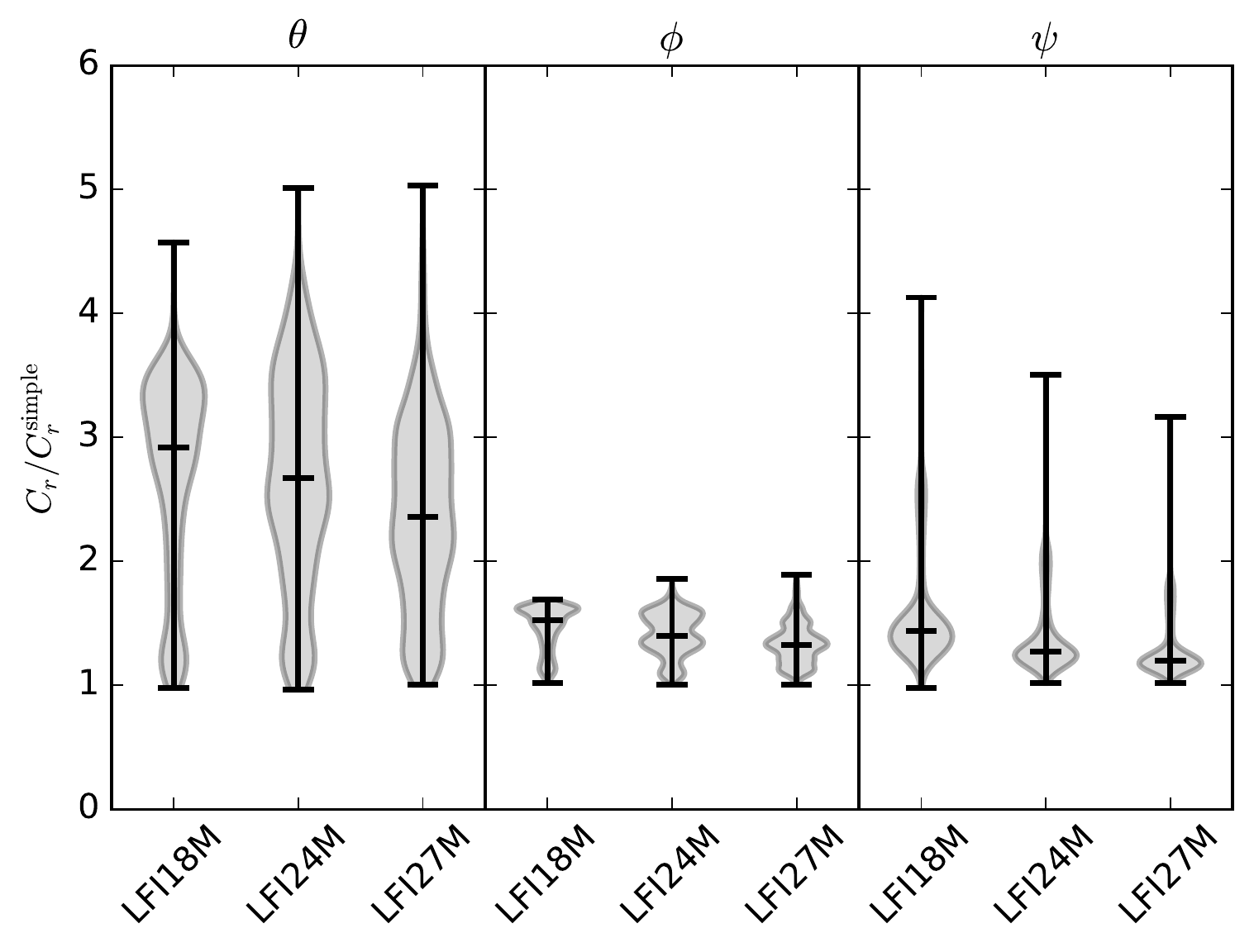}
	\caption{\label{fig:densityPlotsForLFICrs} Improvement in the compression ratio for the LFI TOIs when the full polynomial compression algorithm with the Chebyshev transform step is applied, with respect to the case of the simple compression algorithm described in Sect.~\protect\ref{sec:FitChebyshevRoles}. The nine violin plots show the distribution of the ratio $C_r / C_r^\mathrm{simple}$ for the three datasets $\theta$, $\phi$, and $\psi$, grouped by the three LFI radiometers LFI18M (70\,GHz), LFI24M (44\,GHz), and LFI27M (30\,GHz). The vertical bars show the extrema and the median value. The median values range between 1.20 ($\psi$ dataset, LFI27M) and 2.92 ($\theta$ dataset, LFI18M), showing that the Chebyshev transform step can significantly improve the performance of the compression algorithm.}
\end{figure}

To determine the effectiveness of the Chebyshev transform in improving the compression ratio as compared to the simple algorithm described in Sect.~\ref{sec:FitChebyshevRoles}, I have re-run the optimization for all the data using the simple polynomial compression algorithm. The comparison between the $C_r$ of each best solution in the two cases (polynomial compression versus simple compression) is shown in Fig.~\ref{fig:densityPlotsForLFICrs}. The advantage of the polynomial compression algorithm over the simple algorithm is evident expecially in the three $\theta$ datasets.

I have compressed only three out of 22 LFI radiometers. It is possible to extend this result to forecast the expected compression ratio on all the radiometers. The pointing information of the other 19 radiometers is very similar to one of these, because the \Planck{} focal plane moves rigidly. Therefore, the results obtained for these three radiometers can be generalized to the whole set. Assuming that the overall compression ratio of the three radiometers are representative of all the radiometers with the same sampling frequency, it is easy to derive the following formulae
\begin{align}
N_\mathrm{samples}^\mathrm{in} &= \sum_{f = 30, 44, 70} N_\mathrm{rad}^{(f)}\,\Delta t\,\nu^{(f)}\,n_\mathrm{bits}, \\
N_\mathrm{samples}^\mathrm{out} &= \sum_{f = 30, 44, 70} \frac{N_\mathrm{rad}^{(f)}\,\Delta t\,\nu^{(f)}}{C_r^{(f)}}\,n_\mathrm{bits}, \\
\begin{split}
\label{eq:crForAllLFIRadiometers}
C_r^\mathrm{total} &= \frac{N_\mathrm{samples}^\mathrm{in}}{N_\mathrm{samples}^\mathrm{out}} = \\
&= \frac{\sum_{f = 30, 44, 70} N_\mathrm{rad}^{(f)}\,\nu^{(f)}}{\sum_{f = 30, 44, 70} N_\mathrm{rad}^{(f)}\,\nu^{(f)}/C_r^{(f)}},
\end{split}
\end{align}
where $N_\mathrm{rad}^{(f)}$ is the number of radiometers at frequency $f$, $\nu^{(f)}$ is the sampling frequency, $\,n_\mathrm{bits}$ is the number of bits used to encode each sample (its value is assumed to be the same for the uncompressed and compressed sequences), $\Delta t$ is the acquisition time, roughly equal to four years, and $C_r^{(f)}$ is the compression ratio of the whole FITS file, assumed to be the same for all the $N_\mathrm{rad}^{(f)}$ radiometers. Substituting the values of $C_r^{(f)}$ found for LFI18M, LFI24M, and LFI27M into Eq.~\eqref{eq:crForAllLFIRadiometers} leads to the result
\begin{equation}
C_r^\mathtt{pc} = 9.04,
\end{equation}
as $C_r^{(30)} = 7.40$, $C_r^{(44)} = 8.30$, and $C_r^{(70)} = 9.59$. Since the space needed to keep LFI timelines in uncompressed FITS files is of the order of 7\,TB, this means that compressing such timelines using \libpolycomp{} would produce an archive slightly smaller than 800\,GB.

\section{Impact of the compression in the analysis of \Planck/LFI time series}

\begin{figure*}
	\centering
	\includegraphics[width=\textwidth]{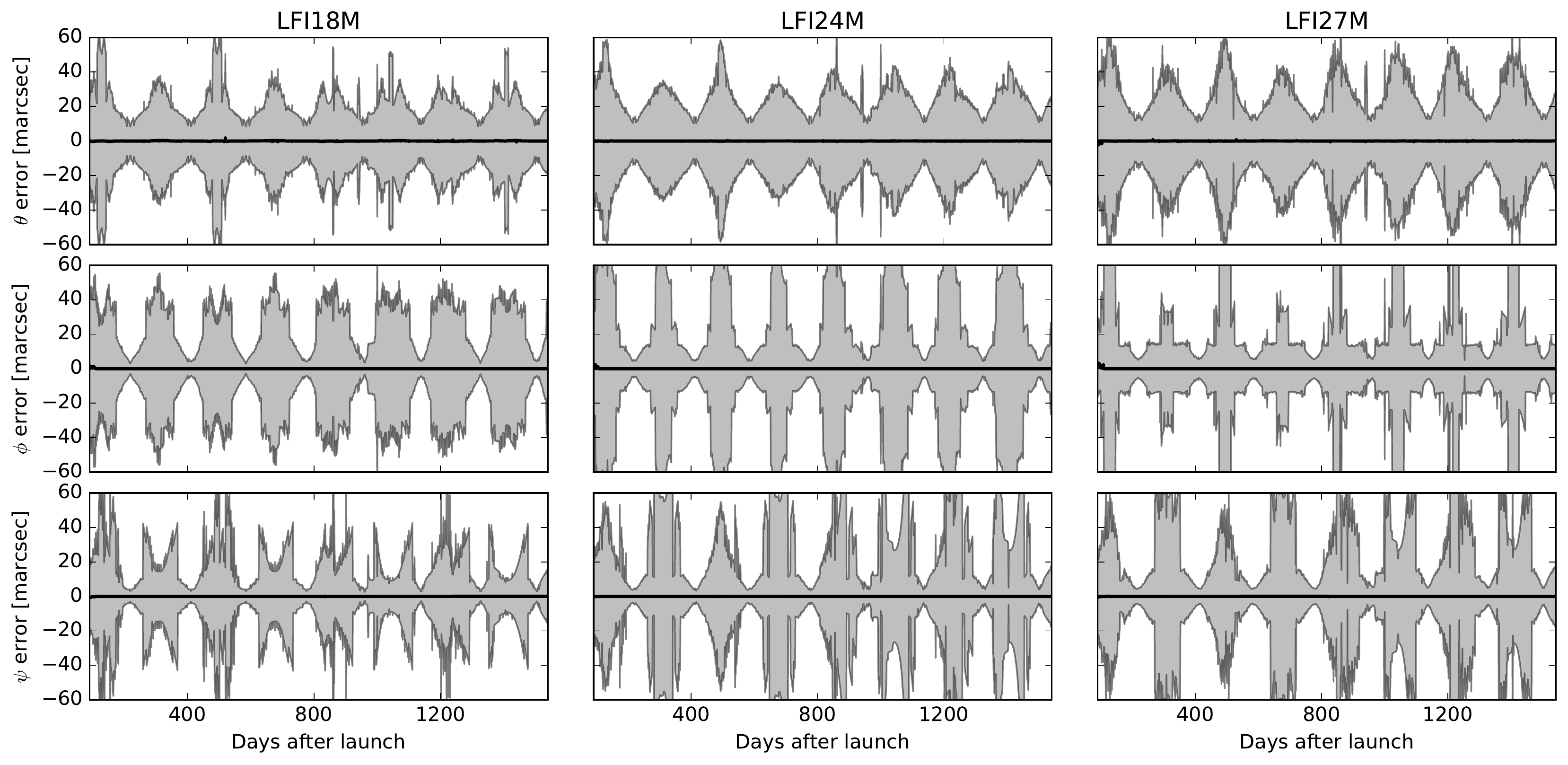}
	\caption{\label{fig:errorsLFI} Compression errors caused by the polynomial compression algorithm when applied to the three pointing angles $\theta$ (colatitude), $\phi$ (longitude), and $\psi$ (orientation) for the LFI TOIs measured by three radiometers, as a function of time. The three radiometer considered are LFI18M (70\,GHz), LFI24M (44\,GHz), LFI27M (30\,GHz). Each plot shows the distribution of the error in each operational day by means of the first, second and third quartiles: the second quartile (median) is represented by the thick line within the filled area, whose boundaries are the first and third quartiles. The minimum and maximum errors are not shown, as they are equal to $\pm \epsilon_c = \pm 1\,\mathrm{arcsec}$ by definition.}
\end{figure*}

In the previous section I analyzed the performance of the compression schemes presented in Sect.~\ref{sec:algorithmDescription}, in terms of the compression ratio $c_R$ (Eq.~\ref{eq:compressionRatio}). However, two other important parameters which quantify the performance of the compression are: (1) the difference between compressed and uncompressed samples, whose upper bound is $\epsilon_c$, defined in Eq.~\ref{eq:compressionError}, and (2) the time required to compress and decompress the data. In this section I discuss the quantification of such parameters in the compression of the LFI TOIs discussed in Sect.~\ref{sec:LFIPointingCompression}.

\subsection{Compression error in the LFI TOI pointing angles}

I discuss here a few statistical properties of the error in the compression of the samples measuring the three pointing angles $\theta$ (Ecliptic colatitude), $\phi$ (Ecliptic latitude), and $\psi$ (orientation) for the three LFI radiometers considered in Sect.~\ref{sec:LFIPointingCompression}.

\begin{figure}
	\centering
	\includegraphics[width=\columnwidth]{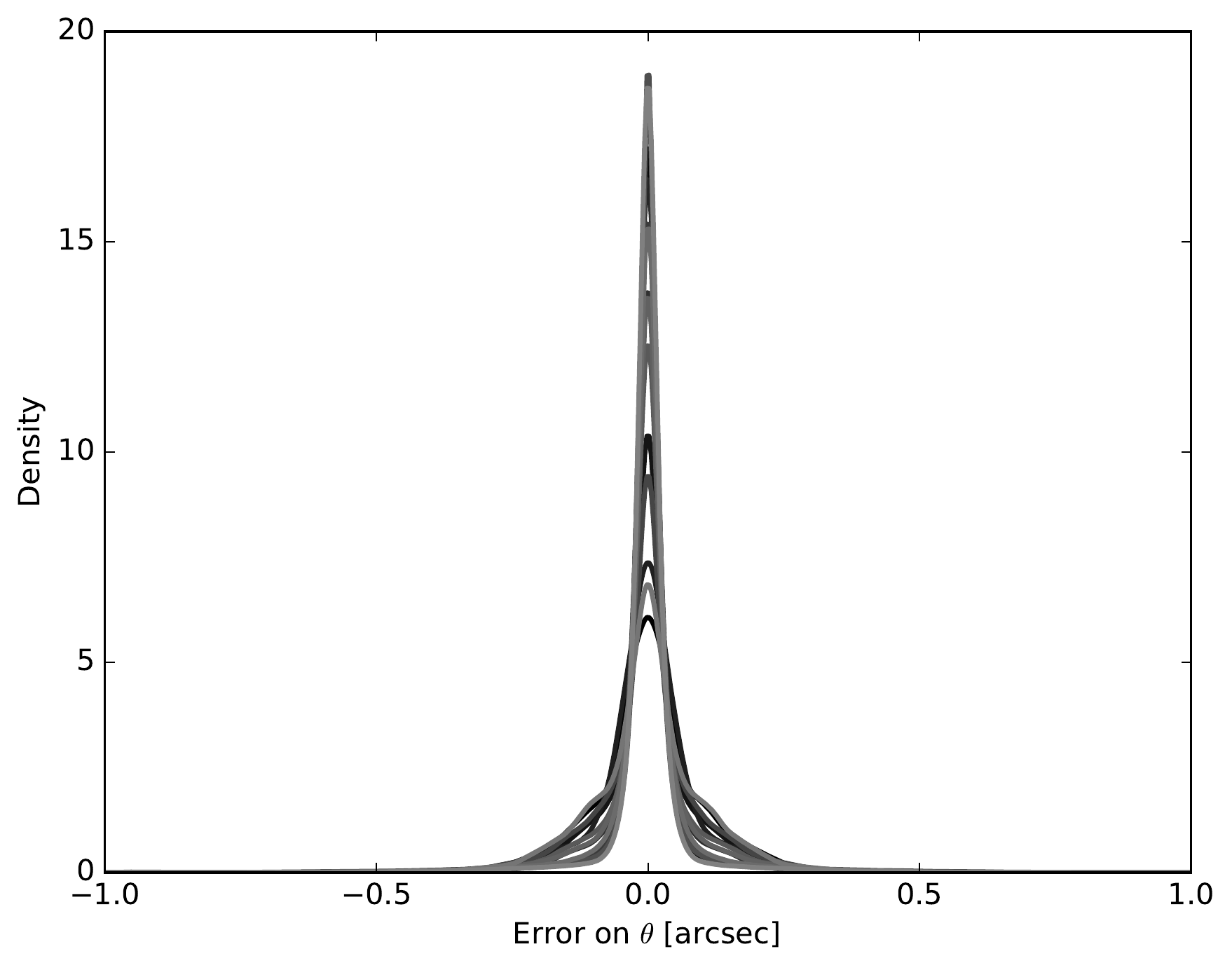}
	\caption{\label{fig:randomOdErrorStats} Distribution of the compression error for the Ecliptic colatitude $\theta$ during each of the fourteen ODs considered in Fig.~\protect\ref{fig:LFIoptimTheta}. The range of the abscissa spans the range $[-\epsilon_c, +\epsilon_c]$. The density has been calculated using the \texttt{stats.gaussian\_kde} function provided by SciPy~0.13.3 (\protect\url{http://www.scipy.org/}). The difference between the mean and median value is of the order of $10^{-5}\,\mathrm{arcsec}$, and the skewness is always less than $(0.5\,\mathrm{arcsec})^3$.}
\end{figure}

My analysis considered the difference between the $j$-th sample $d_j$ and the compressed value $\tilde d_j$:
\begin{equation}
e_j = \tilde d_j - d_j.
\end{equation}
For each of the three radiometers I characterized the statistical properties of each of the 1452 datasets (one per operational day) in terms of the following quantities:
\begin{enumerate}
\item Maximum and minimum value;
\item Median;
\item First and third quantiles.
\end{enumerate}
I computed the maximum and minimum value only as a way to test the correctness of the implementation of the algorithm, as the polynomial compression algorithm ensures that $\left|e_j\right| \leq \epsilon_c$ (\ref{eq:NtDefinitionII}). The values of the median and the quantiles are shown in figure~\ref{fig:errorsLFI}. The average error (median) is consistent with zero in every case, and the inter-quartile range is of the order of tens of milliarcseconds, thus at least one order of magnitude smaller than the upper bound $\epsilon_c = 1\,\mathrm{arcsec}$ set for the compression error of the three LFI pointing angles. The statistical distribution of the errors is not normal, since it is bounded by $\pm \epsilon_c$ by definition; it is sharply peaked around zero, and it shows a remarkable level of simmetry: the difference between the mean error and the median error, as well as the skewness of the error, is of the order of a few tens of arcsecond at most for all the ODs considered in the analysis.

\subsection{Number of hits per pixel in sky maps}
\label{sec:numberOfHits}

\begin{figure}[tb]
    \centering
    \includegraphics[width=\columnwidth]{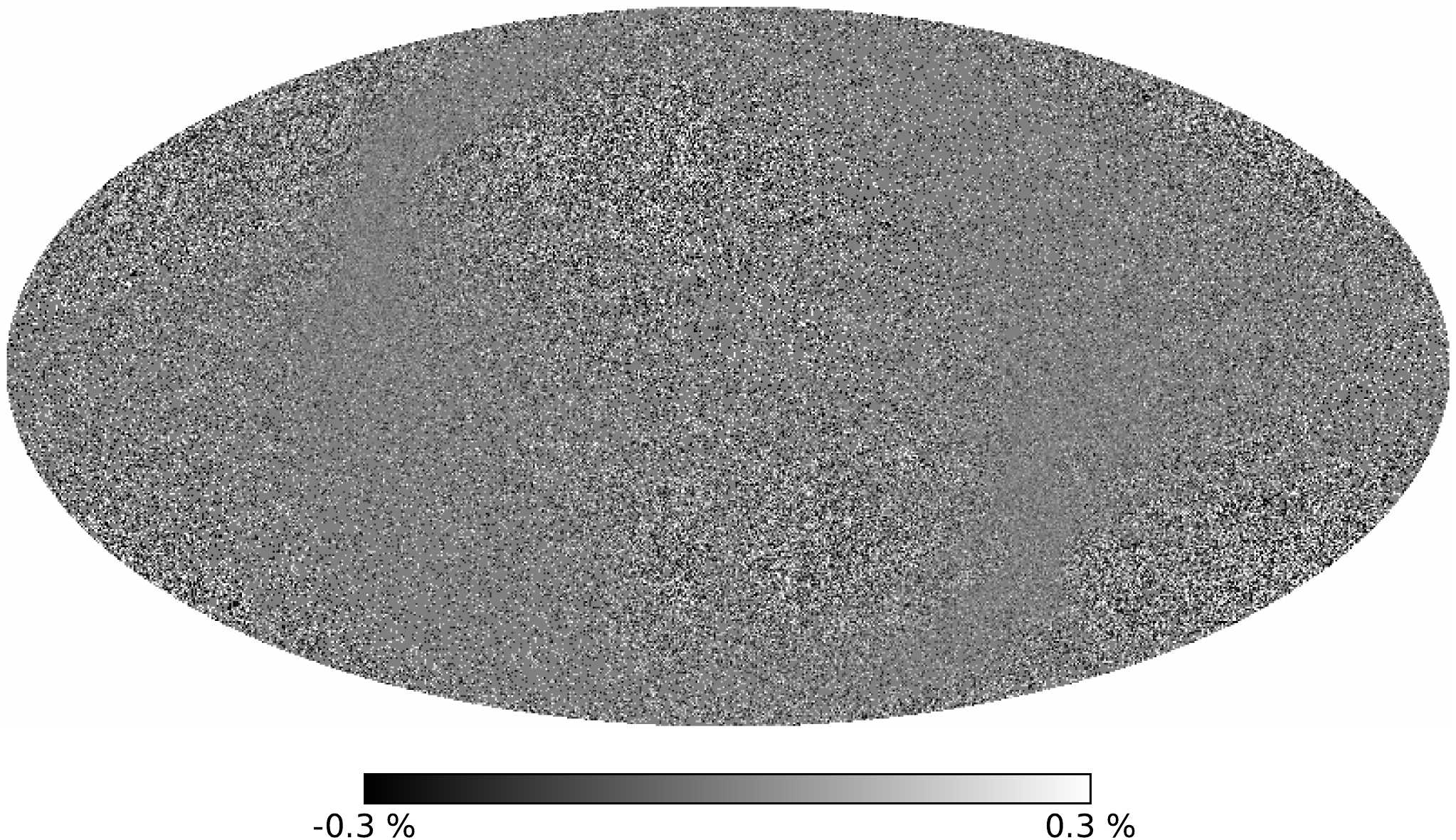}
    \caption{\label{fig:hitMap} \textit{Top:} difference in the number of hits per pixel between a map produced using PLA pointing information for radiometer LFI18M (70\,GHz) and a map produced using the compressed PLA datastreams discussed in Sect.~\protect\ref{sec:LFIPointingCompression}, represented using Ecliptic coordinates. The value of each pixel has been divided by the number of hits. The maximum and minimum pixel values are about $\pm 1.2\,\%$; the color range has been shrunk in order to saturate the colors and better highlight the features of the map. See also Fig.~\protect\ref{fig:hitNumberDistribution}.}
\end{figure}

\begin{figure}[tb]
    \centering
    \includegraphics[width=\columnwidth]{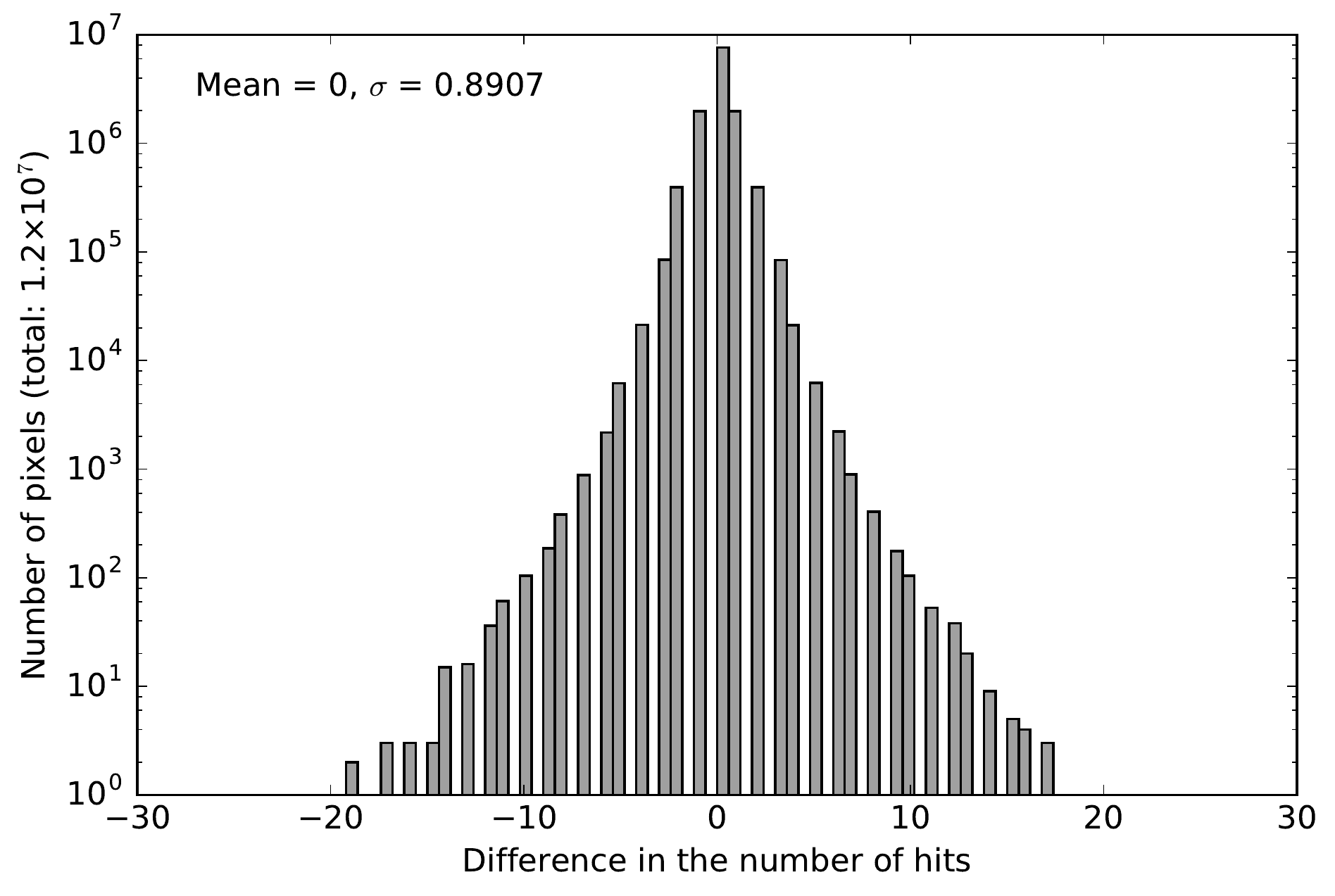}
    \caption{\label{fig:hitNumberDistribution} Distribution of the differences in the hit counts of the two maps used to produce Fig.~\protect\ref{fig:hitMap}. The overall number of hits is $9.8\times 10^9$, and the number of pixels in each map is $12\times N_\text{side}^2 \approx 1.26\times 10^7$ ($N_\text{side} = 1024$). The median of the number of hits is 607, and the minimum and maximum values are 165 and 69\,529.}
\end{figure}

The purpose of measuring timelines using \Planck/LFI is to project each sample on the sky sphere and produce a map of the full sky. Since this process requires the sky sphere to be discretized into a set of pixels \citep[the \Planck{} collaboration uses the Healpix pixelization scheme, see][]{gorski2005}, the value of each pixel will be the combination of the value of one or more samples. The number of samples used to determine the value of a pixel is the \emph{hit count} of the pixel. This quantity has a number of applications (e.g., white noise characterization, destriping), and it is thus interesting to determine if the compression errors in the pointing information alter the hit count significantly.

I have compared hit count maps produced using the original, uncompressed pointings with the same map produced using compressed pointings. Results are shown in Figs.~\ref{fig:hitMap} and \ref{fig:hitNumberDistribution}. Mismatches have zero mean and average, and the overall level of the mismatch is small. The maps use the Healpix pixelization scheme \citep{gorski2005}, with a resolution of $3.4'$ ($N_\text{side} = 1024$), the same resolution as the nominal \Planck{} maps.

\begin{figure}[tbf]
	\centering
	\includegraphics[width=\columnwidth]{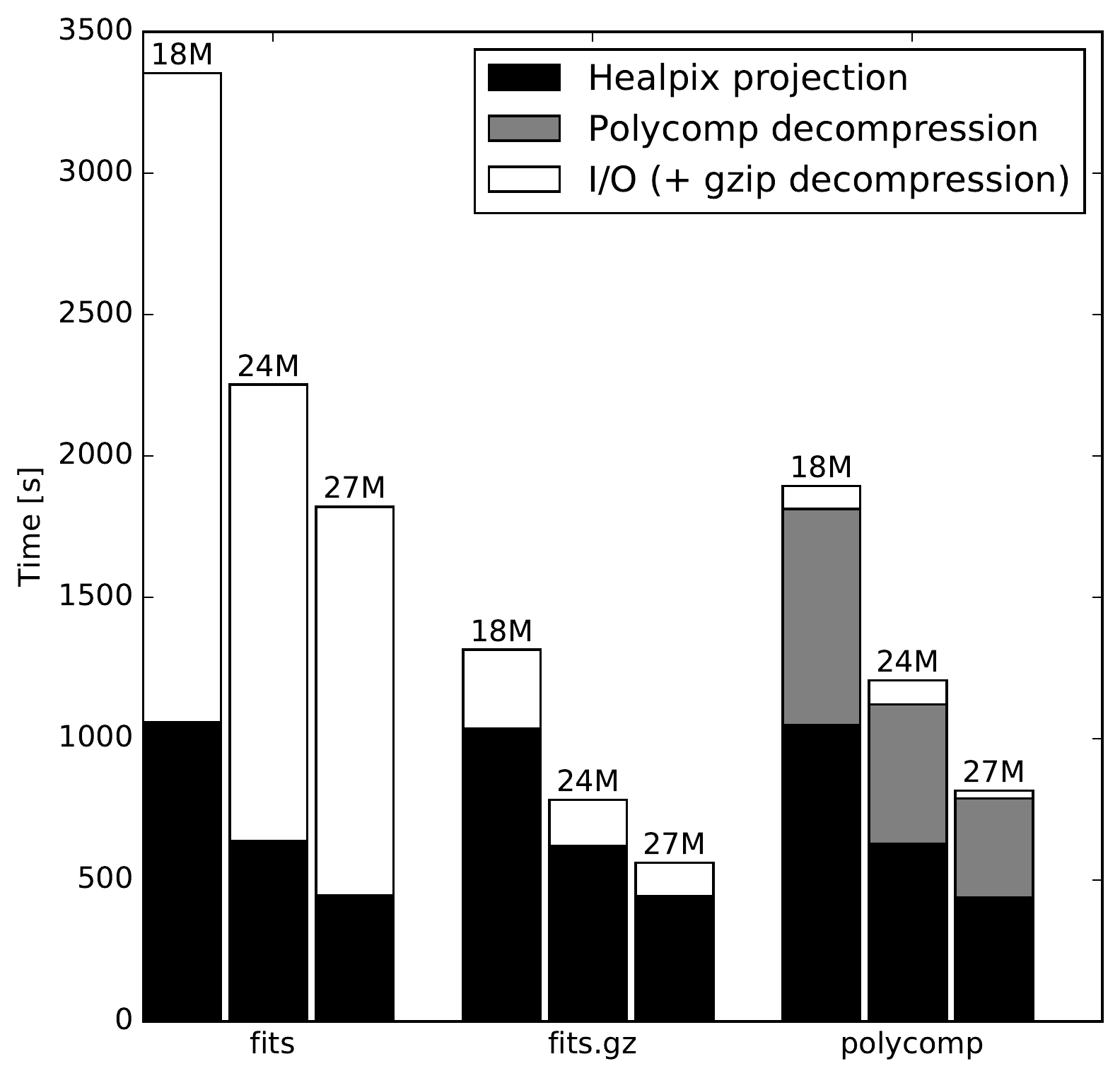}
	\caption{\label{fig:hitMapTimeBarplot} Time required to create hit maps like the one shown in Fig.~\protect\ref{fig:hitMap} on the Erebor cluster. Three ways to store pointing information to disk are shown here: (1) FITS files containing the uncompressed values for $\theta$ and $\phi$; (2) the same as (1), but the files have been compressed using \texttt{gzip}; (3) \polycomp{} files.}
\end{figure}

I have profiled the execution of the script which creates the hit maps, in order to measure the time required to run the following operations:
\begin{enumerate}
\item Loading data from FITS files via calls to \texttt{cfitsio} \citep{ascl.cfitsio};
\item Decompressing data using \libpolycomp{} (when applicable);
\item Projecting the pointings on the sky and creating the map, using my own implementation of Healpix projection functions \citep{gorski2005}.
\end{enumerate}
I run this test on Numenor\footnote{\url{http://www.mi.infn.it/~maino/erebor.html}.}, a 96-core cluster of Intel Xeon processors hosted by the Physics department of the Universit\`a{} degli Studi in Milan. Each run allocated 12 OpenMP processes for \libpolycomp. After having created the hit maps from the pointings stored in \polycomp{} files, I repeated the test twice, reading pointing information that was stored in (1) uncompressed FITS files, (2) \texttt{gzip}-compressed FITS files. Because of the way \texttt{cfitsio} reads data, I did not use \Planck{} PLA files: since FITS binary tables are stored in row-major order and \texttt{cfitsio} disk reads are buffered in chunks of 2880 bytes, reading only a few columns in a PLA file would require \texttt{cfitsio} to load all the six columns in the table HDU from disk. Therefore, I created a new set of FITS files containing only the \texttt{THETA} and \texttt{PHI} columns, and read them in chunks of $N$ rows, where $N$ is the return value of the \texttt{fits\_get\_rowsize} function: this reduces the I/O time for uncompressed FITS files by a factor $\sim 4$ with respect to the case where $\theta$ and $\phi$ are loaded from PLA files via two calls to \texttt{fits\_read\_col}. No such trick is required when the code loads \polycomp{} files, as each column is stored in its own HDU.

The results of the three tests are shown in Fig.~\ref{fig:hitMapTimeBarplot}. Using \polycomp{} files represents a clear advantage over uncompressed FITS files, but the fastest case is when \texttt{gzip}-compressed files are loaded. In this case, \texttt{cfitsio} decompresses the file in memory and no longer accesses the disk. The disadvantage of using gzipped FITS files is that the compression ratio is quite poor: for all the three frequencies, $C_r \approx 1.2$.

\section{Conclusions}

In this paper I have presented the result of three activities:
\begin{enumerate}
\item The description of an algorithm for the compression of smooth data series which approximates data through the sum of least-squares polynomials and Chebyshev polynomials;

\item The implementation of a program, \polycomp, which compresses data series using the algorithm in point~1 as well as other well-known compression algorithms;

\item A characterization of \polycomp's ability to compress \Planck/LFI time ordered information, both in terms of the compression ratio (Eqs.~\ref{eq:compressionRatio} and \ref{eq:compressionRatioPc}) and of the compression error (Eq.~\ref{eq:compressionError}). Given some reasonable upper bound to the compression error, the achieved compression ratio is greater than 8. I have also estimated the impact of the compression error on a few quantities relevant for the analysis of the \Planck/LFI data.
\end{enumerate}

The results presented in this paper might find interesting application in the development of techniques for the storage of data acquired by future space missions. An example is the proposed \textit{LiteBird} mission \citep{matsumura.2014.litebird}, which will be devoted to the measurement of CMB polarization anisotropies in the 50-320\,GHz range: the instrument will be made by 100 times many sensors as \Planck/LFI and is therefore likely have significant demand in terms of data storage.

\section*{Aknowledgements}
\label{sec::aknowledgements}

The author would like to thank Michele Maris for having introduced him into the world of astronomical data compression, and Marco Bersanelli and the two anonymous referees for the useful comments which helped to improve this paper and the \libpolycomp{} library.

\section*{References}

\bibliographystyle{model2-names}\biboptions{authoryear}
\bibliography{polycomp}

\appendix

\section{Implementation of {\tt polycomp}}
\label{sec:Polycomp}

I have implemented the algorithms described in this paper in a BSD-licensed C library, \libpolycomp\footnote{\url{https://github.com/ziotom78/libpolycomp}.}. I have also implemented a set of Python~3 bindings to the library, available in a separate repository (\url{https://github.com/ziotom78/polycomp}). The Python library includes a stand-alone program, \polycomp{}, which can compress/decompress ASCII and binary tables saved in FITS files.

The \libpolycomp{} library has been implemented using the 1989 definition of the C standard, and it should therefore be easily portable to different compilers. The author tested it using the following compilers:
\begin{itemize}
\item GNU \texttt{gcc}\footnote{\url{https://gcc.gnu.org/}.} 4.9 and 5.1;
\item \texttt{clang}\footnote{\url{http://clang.llvm.org/}.} 3.4 and 3.5;
\item Intel C Compiler\footnote{\url{https://software.intel.com/en-us/c-compilers}.} 16.0.
\end{itemize}
The library uses OpenMP \citep{OpenMPSpec} to take advantage of multiple-core systems. It has been fully documented, and the user's manual\footnote{\url{http://ziotom78.github.io/libpolycomp/}.} is available online. The library API has been designed in order to be easily callable from other languages.

The Python wrappers have been built using Cython\footnote{\url{http://cython.org/}.}, and the \polycomp{} program is able to save the compressed timestreams in files. The format of these files is based on the FITS file format \citep{Pence2010}, and it is fully documented in the \polycomp{} user's manual\footnote{\url{http://polycomp.readthedocs.org/en/latest/}.}.

\end{document}